\documentclass[11pt]{amsart}
\usepackage{amssymb, latexsym}

\newcommand{\beqa}{\begin{eqnarray*}}
\newcommand{\eeqa}{\end{eqnarray*}}
\newcommand{\beqn}{\begin{eqnarray}}
\newcommand{\eeqn}{\end{eqnarray}}

\newcommand{\iy}{\infty}
\newcommand{\lt}{\left}
\newcommand{\rt}{\right}

\newcommand{\R}{\mathbb R}

\newcommand{\N}{\mathbb N}

\newcommand{\mcH}{\mathcal H}
\newcommand{\mcB}{\mathcal B}
\newcommand{\mcC}{\mathcal C}

\newcommand{\al}{\alpha}

\newcommand{\la}{\lambda}

\newcommand{\Om}{\Omega}

\newcommand{\s}{\sigma}
\newcounter{cnt1}
\newcounter{cnt2}
\newcounter{cnt3}
\newcommand{\blr}{\begin{list}{$($\roman{cnt1}$)$}
 {\usecounter{cnt1} \setlength{\topsep}{0pt}
 \setlength{\itemsep}{0pt}}}
\newcommand{\bla}{\begin{list}{$($\alph{cnt2}$)$}
 {\usecounter{cnt2} \setlength{\topsep}{0pt}
 \setlength{\itemsep}{0pt}}}
\newcommand{\bln}{\begin{list}{$($\arabic{cnt3}$)$}
 {\usecounter{cnt3} \setlength{\topsep}{0pt}
 \setlength{\itemsep}{0pt}}}
\newcommand{\el}{\end{list}}
\newtheorem{thm}{Theorem}[section]
\newtheorem{lem}[thm]{Lemma}
\newtheorem{cor}[thm]{Corollary}
\newtheorem{ex}[thm]{Example}

\newtheorem{Def}[thm]{Definition}

\newtheorem{rem}[thm]{Remark}
\newcommand{\Rem}{\begin{rem} \rm}
\newcommand{\bdfn}{\begin{Def} \rm}
\newcommand{\edfn}{\end{Def}}

\newcommand{\ba}{\begin{array}}
\newcommand{\ea}{\end{array}}

\newtheorem*{acknowledgements}{Acknowledgements}
\numberwithin{equation}{section}
\date{}
\begin{document}
\title{\bf  Adjoint Operators on Banach Spaces}
\author[Gill]{T. L. Gill}
\address[Tepper L. Gill]{ Departments of Mathematics, Physics, and Electrical \& Computer Engineering, Howard University\\
Washington DC 20059 \\ USA, {\it E-mail~:} {\tt tgill@howard.edu}}
\author[Mensah]{F.  Mensah}
\address[Francis Mensah]{ Department of Mathematics \\ Howard
University\\ Washington DC 20059 \\ USA, {\it E-mail~:} {\tt
mensah3@yahoo.com}}
\author[Zachary]{W. W. Zachary}
\address[Woodford W. Zachary]{ Departments of Mathematics, and Electrical \& Computer Engineering, \\ Howard
University\\ Washington DC 20059 \\ USA, {\it E-mail~:} {\tt
wwzachary@earthlink.net}}
\date{}
\maketitle
\begin{abstract}    In this paper, we report on new results related to the existence of an adjoint for operators on separable Banach spaces and discuss a few interesting applications.  (Some results are new even for Hilbert spaces.)  Our first two applications provide an extension of the  Poincar\'{e} inequality and the Stone-von Neumann version of the spectral theorem for a large class of $C_0$-generators of contraction semigroups on separable Banach spaces.  Our third application provides a natural extension of the Schatten-class of operators to all separable Banach spaces.  As a part of this program, we introduce a new class of separable Banach spaces.  As a side benefit, these spaces also provide a natural framework for the (rigorous) construction of the path integral as envisioned by Feynman.       
\end{abstract}

\tableofcontents
\section{\bf{Adjoint Theory}}
\subsection{Introduction}
One of the impediments to the development of a clear parallel
theory for operators on Banach spaces compared to that for Hilbert spaces is the lack of a suitable notion of an adjoint operator.   In this section we use a Theorem of Gross and Kuelbs to construct an adjoint for all bounded linear operators on a separable Banach space. We then show that this result can be extended to all closed densely defined linear operators of Baire class one (limits of bounded linear operators). We use these results in later sections to extend the Poincar\'{e} inequality, the spectral theorem and to construct the ``natural" Banach space version of the Schatten class.
\subsection{Background}
Let  ${\mathcal{B}}$ be a separable Banach space over the complex field and let $L[{\mathcal{B}}]$ denote the bounded linear operators on  ${\mathcal{B}}$.  Assume that  ${\mathcal{B}}$ has a continuous dense embedding in a Hilbert space ${\mathcal{H}}$.  By a duality map, $f_u: \mcB \rightarrow {\mcB}'$,  we mean any linear functional $f_{u} \in \{f \in \mathcal{B'}| \ f(u)=<u,f> = \|u\|^{2}_\mathcal{B}=\|f\|^{2}_{\mathcal{B}'}, \ u \in \mathcal{B}\}$,
where $<.>$ is the natural pairing between a Banach space and its dual. Let $\bf{J}: \mathcal{H} \longrightarrow \mathcal{H'}$ be the standard conjugate isomorphism between a Hilbert space and its dual, so that ${<u, {\bf J}(u)>} = (u,u)_\mathcal{H}=\|u\|^{2}_{\mathcal{H}}$.

For fixed $u$ define a seminorm $p_u ( \ \cdot \ )$ on $\mcB$ by  $p_u (x) = \lt\| u \rt\|_{\mcB} \lt\| x \rt\|_{\mcB}$,  and define $ {\hat f}_u^s (\, \cdot \,) $ by:
\[
{\hat f}_u^s (x) = \frac{{\left\| u \right\|_{\mcB}^2 }}
{{\left\| u \right\|_{\mcH}^2 }}\left( {x,u} \right)_{\mcH} .
\]
On the closed subspace $M= \left\langle {u} \right\rangle, \; \left| {{\hat f}_u^s (x)} \right| = \left\| u \right\|_B \left\| x \right\|_B  \leqslant p_u (x)$.  By the (complex version of the) Hahn-Banach Theorem, ${\hat f}_u^s (\, \cdot \,)$ has an extension, $f_u^s (\, \cdot \,)$, to $\mcB$ such that $\left| {f_u^s (x)} \right|   \leqslant p_u (x)= \left\| u \right\|_B \left\| x \right\|_B$ for all $x \in \mcB$ (see Rudin \cite{RU}, Theorem 3.3, page 57). From here, we see that $\left\| {f_u^s } \right\|_{\mathcal{B}'} \le \left\| u \right\|_\mathcal{B}$.

On the other hand, we have:
\[
f_u^s (u) = \left\| u \right\|_\mathcal{B}^2  \leqslant \left\| u \right\|_\mathcal{B} \left\| {f_u^s } \right\|_{\mathcal{B}'}  \Rightarrow \left\| u \right\|_\mathcal{B}  \leqslant \left\| {f_u^s } \right\|_{\mathcal{B}'}, 
\]
so that $f_u^s ( \, \cdot \, )$ is a duality mapping for $u$.  We call $f_u^s ( \, \cdot \, )$ the {\it Steadman duality map} on $\mathcal{B}$ associated with $\mathcal{H}$. 

Recall that a densely defined operator  ${A}$ is called accretive if $\operatorname{Re} \left\langle {Au ,f_u^s} \right\rangle  \ge 0$ for $u \in D(A)$; and  it is called m-accretive if, in addition, it is closed and $Ran(I+A)=\mathcal{B}$.  The following theorem by Lax \cite{L} is important for our theory.  It is not as well-known as it should be, so we provide a proof of the first part.  We prove a stronger version of parts two and three in Section 3 (see Theorem 3.3, part 2).
\begin{thm}[Lax]\label{L: lax}  Suppose ${\mathcal{B}}$ is a dense continuous embedding in a separable Hilbert space 
${\mathcal{H}}$.   Let $A \in L[{\mathcal{B}}]$.  If $A$ is selfadjoint on ${\mathcal{H}}$
(i.e., $\left( {Ax,y} \right)_{\mathcal{H}}  = \left( {x,Ay} \right)_{\mathcal{H}}, 
\forall x{\text{,}}y \in {\mathcal{B}}$), then 
\begin{enumerate}
\item The operator $A$ is bounded on ${\mathcal{H}}$ and $
\left\| A \right\|_{\mathcal{H}}  \leqslant k \left\| A \right\|_{\mathcal{B}}$, for some positive constant $k$.
\item The spectra of $A$ over $\mathcal{H}$ and over $\mathcal{B}$ satisfy $\sigma _\mathcal{H} (A) \subset \sigma _\mathcal{B} (A)$.
\item The point spectrum of $A$ is unchanged by the extension (i.e., $\sigma _\mathcal{H}^p (A) =  \sigma _\mathcal{B}^p (A)$).
\end{enumerate}
\end{thm}
\begin{proof} To prove (1), let $x  \in {\mathcal{B}}$
 and, without loss, we can assume that $k = 1$ and $\left\| x  \right\|_\mathcal{H}  = 1$.  Since $A$ is selfadjoint, 
\[
\left\| {Ax } \right\|_\mathcal{H}^2  = \left( {Ax ,Ax } \right) = \left( {x ,A^2 x } \right) \leqslant \left\| x  \right\|_\mathcal{H} \left\| {A^2 x } \right\|_\mathcal{H}  = \left\| {A^2 x } \right\|_\mathcal{H}. 
\]
Thus, we have $\left\| {Ax } \right\|_\mathcal{H}^4  \leqslant \left\| {A^4 x } \right\|_\mathcal{H}$, so it is easy to see that $
\left\| {Ax } \right\|_\mathcal{H}^{2n}  \leqslant \left\| {A^{2n} x } \right\|_\mathcal{H}$ for all $n$.  It follows that: 
\[
\begin{gathered}
  \left\| {Ax } \right\|_\mathcal{H}  \leqslant (\left\| {A^{2n} x } \right\|_\mathcal{H} )^{1/2n}  \leqslant (\left\| {A^{2n} x } \right\|_\mathcal{B} )^{1/2n}  \hfill \\
  {\text{          }} \leqslant (\left\| {A^{2n} } \right\|_\mathcal{B} )^{1/2n} (\left\| x  \right\|_\mathcal{B} )^{1/2n}  \leqslant \left\| A \right\|_\mathcal{B} (\left\| x  \right\|_\mathcal{B} )^{1/2n}.  \hfill \\ 
\end{gathered} 
\]
Letting $n \to \infty $, we get that $\left\| {Ax } \right\|_\mathcal{H}  \leqslant \left\| A \right\|_{\mathcal{B}} $ for $x $ in a dense set of the unit ball of $
\mathcal{H}$.  We are done, since the norm is attained on a dense set of the unit ball.
\end{proof}
The following is a result due to Gross and Kuelbs [GR], [KB].  \begin{thm} \label{GK} Suppose ${\mathcal{B}}$ is a separable Banach space.  Then there exist separable Hilbert spaces ${\mathcal{H}}_1 ,{\mathcal{H}}_2$ and a positive trace class operator ${\mathbf{T}}_{12}$ defined on ${\mathcal{H}}_2$ such that ${\mathcal{H}}_1  \subset {\mathcal{B}} \subset {\mathcal{H}}_2 $ (all as continuous dense embeddings).
\end{thm}
\begin{proof} As ${\mathcal{B}}$ is separable, let $\{ u_n \} $ be a dense set in ${\mathcal{B}}$  and let $\{ f_n \} $ be any fixed set of corresponding duality mappings (i.e., $f_n  \in {\mathcal{B'}}$ and $
f_n (u_n ) = \left\langle {u_n ,f_n } \right\rangle  = \left\| {u_n } \right\|_{\mathcal{B}}^2 = \left\| {f_n } \right\|_{\mathcal{B}'}^2 $).   Let $\{ t_n \}$ be a positive sequence of numbers such that $\sum\nolimits_{n = 1}^\infty  {t_n }  = 1$, and define $\left( {u,v} \right)_2$ by:
\[
\left( {u,v} \right)_2  = \sum\nolimits_{n = 1}^\infty  {t_n f_n (u)} \bar f_n (v).
\]
It is easy to see that $\left( {u,v} \right)_2 $ is an inner product on ${\mathcal{B}}$.  We let $
{\mathcal{H}}_2 $ be the Hilbert space generated by the completion of ${\mathcal{B}}$ with respect to this inner product.  It is clear that ${\mathcal{B}}$ is dense in ${\mathcal{H}}_2 $, and as
\[
\left\| u \right\|_2^2  = \sum\nolimits_{n = 1}^\infty  {t_n \left| {f_n (u)} \right|^2 }  \le \sup _n \left| {f_n (u)} \right|^2  = \left\| u \right\|_{\mathcal{B}}^2,
\]
we see that the embedding is continuous.
\paragraph{}
Now, let  $\{ \varphi _n \}  \in \mcB$ be a complete orthonormal sequence for $
{\mathcal{H}}_2 $, and let $\{ \lambda _n \}$ be a positive sequence such that $
\sum\nolimits_{n = 1}^\infty  {\lambda _n }  < \infty $, and $M = \sum\nolimits_{n = 1}^\infty  {\lambda _n^2 } \left\| {\varphi _n } \right\|_{\mathcal{B}}^2  < \infty $.  Define the operator ${\mathbf{T}}_{12}$ on ${\mathcal{B}}$ by:
\[
{\mathbf{T}}_{12} u = \sum\nolimits_{n = 1}^\infty  {\lambda _n } \left( {u,\varphi _n } \right)_2 \varphi _n.
\]
Since ${\mathcal{B}} \subset {\mathcal{H}}_2,\,   \left( { u ,\varphi _n } \right)_2$ is defined for all $u \in {\mathcal{B}}$.  Thus,  ${\mathbf{T}}_{12}$ maps ${\mathcal{B}} \to {\mathcal{B}}$ and:
\[
\left\| {{\mathbf{T}}_{12} u} \right\|_{\mathcal{B}}^2  \le \left[ {\sum\nolimits_{n = 1}^\infty  {\lambda _n^2 } \left\| {\varphi _n } \right\|_{\mathcal{B}}^2 } \right]\left[ {\sum\nolimits_{n = 1}^\infty  {\left| {\left( {u,\varphi _n } \right)_2 } \right|^2 } } \right] = M\left\| u \right\|_2^2  \le M\left\| u \right\|_{\mathcal{B}}^2.
\]
Thus, ${\mathbf{T}}_{12}$ is a bounded operator on ${\mathcal{B}}$.  Define $ {\mathcal{H}}_1$ by:
\[
{\mathcal{H}}_1  = \left\{ {u \in \mcB \,\left| {\;\sum\nolimits_{n = 1}^\infty  {\lambda _n^{ - 1} \left| {\left( {u,\varphi _n } \right)_2 } \right|} ^2  < \infty } \right.} \right\},\quad \left( {u,v} \right)_1  = \sum\nolimits_{n = 1}^\infty  {\lambda _n^{ - 1} \left( {u,\varphi _n } \right)_2 } \left( {\varphi _n ,v} \right)_2.
\]
With the above inner product, ${\mathcal{H}}_1 $ is a Hilbert space and, since terms of the form $\{u_N  = \sum\nolimits_{k = 1}^N {\lambda _k^{ - 1} } \left( {u,\psi _k } \right)_2 \varphi _k :\;u,\,\psi _k \in \mcB \}$ are dense in ${\mathcal{B}}$, we see that ${\mathcal{H}}_1 $ is dense in $
{\mathcal{B}}$.  It follows that ${\mathcal{H}}_1$ is also dense in ${\mathcal{H}}_2 $.  It is easy to see that ${\mathbf{T}}_{12} $ is a positive selfadjoint operator with respect to the $
{\mathcal{H}}_2$ inner product so, by Theorem 1.1, ${\mathbf{T}}_{12}$ has a bounded extension to ${\mathcal{H}}_2$ and
$
\left\| {{\mathbf{T}}_{12} } \right\|_2  \le \left\| {{\mathbf{T}}_{12} } \right\|_{\mathcal{B}}. $
 Finally, it is easy to see that, for $u,v \in {\mathcal{H}}_1$, $
(u,v)_1  = ({\mathbf{T}}_{12}^{ - 1/2} u,{\mathbf{T}}_{12}^{ - 1/2} v)_2$ and $
(u,v)_2  = ({\mathbf{T}}_{12}^{1/2} u,{\mathbf{T}}_{12}^{1/2} v)_1$.   It follows that $
{\mathcal{H}}_1$ is continuously embedded in ${\mathcal{H}}_2$, hence also in ${\mathcal{B}}$.  
\end{proof}

The construction of ${\mathcal{H}}_1$ and ${\mathcal{H}}_2$ is not unique.  In the next section, we construct a concrete version of ${\mathcal{H}}_2$ which is unique in the sense that we use a fixed dense family $\{ u_n \} \subset \mcB$, a fixed family of linear functionals  $\{ F_n \} \in  {\mcB}'$ and a fixed family of positive numbers $\{ t_n \}$. (We will discuss this more in the remarks before Section 2.1.)    For the remainder of this paper, we assume that both  ${\mathcal{H}}_1$ and  ${\mathcal{H}}_2$ are fixed.
\subsection{Banach Space Adjoint}
The following is the major result in Gill et al [GBZS].   It generalizes the well-known result of von Neumann [VN] for bounded operators on Hilbert spaces.  For convenience, we provide a proof. (We delay the proof of (1) and (3) until after Theorem 1.4.) 
\begin{thm}\label{V: von} Let $A$ be a bounded linear operator on ${\mathcal{B}}$.  Then $A$ has a well-defined adjoint $A^ *$  defined on ${\mathcal{B}}$ such that:
\begin{enumerate}
\item the operator $
A^ * A \ge 0$  (maximal accretive),		
\item $(A^ * A)^ *   = A^ * A$ (selfadjoint), and 
\item $I + A^ * A$ has a bounded inverse.
\end{enumerate}
\end{thm}
\begin{proof}   If we let ${\bf J}_i :{\mcH}_i  \to {\mcH}'_i $, $(i =1, 2)$,
then
${A}_1   =  { A}_{|{\mcH}_{1}  }: {\mcH}_1
\rightarrow {\mcH}_2 ,$ and ${{ A'}_1 :{\mcH}'_2  \rightarrow {\mcH}'_1}.$
 
It follows  that ${ A'}_1 {\bf J}_2 :{{\mcH}_2}  \rightarrow {\mcH}'_1 $
and ${\bf J}_1^{ - 1} {A'}_1 {\bf J}_2 :{\mcH}_2  \to {\mcH}_1  \subset
{\mcB}$ so that, if we define
${A}^ {*}   = [ {{\bf J}_1^{ - 1} {A'}_1 {\bf J}_2 } ]_{\mcB},
$ then ${ A}^ *  :{\mcB} \to {\mcB}$ (i.e., ${A}^ *   \in L[{\mcB}]$).

To prove (2),   we have that  for
$x \in {\mcH}_1,$
\beqa ({A}^ *{A})^ *  x&  = & ( {\{ {{\bf J}_1^{ - 1} [ {\{ {[ {{\bf
J}_1^{  - 1} {A'}_1 {\bf J}_2 } ]| {_{\mcB} }{A}} \}_1 } ]^\prime
{\bf  J}_2 } \}| {_{\mcB} } } )x  \\
  &  = &( {\{ {{\bf J}_1^{ - 1} [ {\{ {{A'}_1 [ {{\bf J}_2 {A}_1 {\bf
J}_1^{ - 1} } ]| {_{\mcB} } } \}} ]{\bf J}_2 } \}|  {_{\mcB} } } )x \\ &=
&{A}^ *{A}x.
\eeqa
It follows that the same result holds on ${\mcB}.$
\end{proof}
The operator  $A^*A$ is selfadjoint on $\mathcal{B}$.   By Theorem 1.1 (of Lax \cite{L}), it is natural to expect that the same is true on ${\mathcal{ H}_{2}}$.  However, this need not be the case.  To obtain a simple counterexample, recall that, in standard notation, the simplest class of bounded linear operators on $\mathcal{B}$ is ${\mathcal{B}}\otimes {\mathcal{B}'}$, in the sense that: 
\beqa
 {\mathcal{B}}\otimes {\mathcal{B}'}: {\mathcal{B}} \to {\mathcal{B}}, \; {\text{by}}\;  Au=(b \otimes  l_{b'}(\cdot))u = \left\langle {b',u} \right\rangle b.
\eeqa
Thus, if $l_{b'}(\cdot) \in {\mathcal{B'}} \backslash {\mathcal{H}}_2^{'}$, then $J_2\{J_1^{-1}[(A_1)']J_2|_{\mathcal{B}}(u)\}$ is not in ${\mathcal{H}}_2^{'}$, so that $A^*A$ is not defined as an operator on all of ${\mathcal{ H}_{2}}$ and thus cannot have a bounded extension. 
 
We can now state the correct extension of Theorem 1.1. (This result corrects an error in \cite{GBZS}.)
\begin{thm}\label{L*: lax*} Let $A$ be a bounded linear operator on ${\mathcal{B}}$.   If ${\mathcal{B}^{'}}\subset {\mathcal{H}}_2$, then $A$ has a bounded extension to $L[{\mathcal{H}_{2}} ]$, with $
\left\| A \right\|_{\mathcal{H}_{2}}  \le k\left\| A \right\|_\mathcal{B}$ (for some positive $k$).  
\end{thm}
\begin{proof}  If $T=A^{*}A$, under the stated conditions, then $\left\langle {Tx,{\mathbf{J}}_2 (y)} \right\rangle  = \left( {Tx,y} \right)_{H_2 }$ is well defined for all $x, \ y \in \mcB$, and $\left( {Tx,y} \right)_{H_2 }  = \left( {x,Ty} \right)_{H_2 }$.  Thus,  we can now apply Lax's Theorem to see that
$\lt\|T\rt\|_{{\mcH}_2}=\lt\|A\rt\|_{{\mcH}_2}^2 \le k^2\lt\|A\rt\|_{\mcB}^2$.  
\end{proof}
We can now finish our proof of Theorem 1.3.

To prove (1), let $x \in {\mcB}$, then $
\lt( {{A}^ *{A}x, x} \rt)_{{\mcH}_2}  \ge 0 $ for all $x \in \mcB$.  Hence 
$\langle {{A}^ * {A}x, f _x^s } \rangle    \geq 0$, so that $A^*A$ is maximal accretive.
The proof of (3), that  ${I} + {A}^ *{A}$ is invertible, follows the same lines as in von Neumann's theorem.
\begin{rem}
Theorem 1.4 tells us that $L[{\mcB}] \subset L[{\mcH}_2]$ as a continuous embedding. (In section 6, we will show that if $\mcB$ has the approximation propertiy, the embedding is dense.) 

The algebra $L[{\mcB}]$ also has a $*$-operation that makes it much closer to $L[{\mcH}_2]$ then expected.  However, in general $\left\| {A^* A} \right\|_{\mcB} \ne \left\| A \right\|_{\mcB}^2 $.  Furthermore, if $A \ne B, \, B^*$ then,  unless
\[
\left( {B\left| {_{{\mcH}_1 } } \right.} \right)^\prime  \left( {A\left| {_{{\mcH}_1 } } \right.} \right)^\prime   = \left( {AB\left| {_{{\mcH}_1 } } \right.} \right)^\prime  ,\quad \left( {AB} \right)^*  \ne A^* B^* .
\]  
Thus, $L[{\mcB}]$ is a not a $*$-algebra in the traditional sense.  
\end{rem}
\section{\bf {The Hilbert Space} ${\bf{KS}}^2$}   
Theorem 1.4 is odd, given the requirement that ${\mcB}' \subset {\mcH}_2$.  The following example shows that, if it's true at all, it does not work for one of the standard Banach-Hilbert space couples.  
\begin{ex}
Let $\ell_1\to \ell_2$ be the natural embedding, and let $e_n$ be the natural unit basis. Put $T(e_1)=e_1$ and $T(e_n)=e_1+e_n$ for $n>1$.
This operator has a natural extension to a bounded linear operator in $\ell_1$. Put $x_n=n^{-1}(e_1+\dots+e_n)$.   Then $\lt\|x_n\rt\|_2 \to  0$, $\lt\| T(x_n)-e_1\rt\|_2 \to  0$ but $T(0)\neq e_1$.  Thus, $T$ cannot be extended to a closed operator on  $
\ell_2$.  It follows that $\ell_2$ is not the correct Hilbert space for the extension of bounded linear operators or for the construction of adjoints for bounded linear operators on $\ell_1$.  (Note that ${\ell_1}'$ is not contained in ${\ell_2}$.)
\end{ex}
The purpose of this section is to construct a Hilbert space which allows us to apply Theorem 1.4 to all classical Banach spaces. 

In order to construct the space of interest, first recall that Alexiewicz \cite{AL} has shown that the class $D({\R})$, of Denjoy integrable functions (restricted and wide sense), can be normed in the following manner:  for $f \in D({\R})$, define $\left\| f \right\|_D$ by
\beqn
\left\| f \right\|_D  = \sup _s \left| {\int_{ - \infty }^s {f(r)dr} } \right|.
\eeqn
It is clear that this is a norm, and it is known that $D({\R})$ is not complete.   Replacing ${\mathbb{R}}$ by ${\mathbb{R}}^n$ in (2.1), for $f \in D({\mathbb{R}}^n )$, we have:
\beqn
\left\| f \right\|_D  = \sup _{r > 0} \left| {\int_{{\mathbf{B}}_r } {f({\mathbf{x}})d{\mathbf{x}}} } \right| = \sup _{r > 0} \left| {\int_{\mathbf{R}^n } {{\mathcal{E}}_{{\mathbf{B}}_r } ({\mathbf{x}})f({\mathbf{x}})d{\mathbf{x}}} } \right| < \infty, 
\eeqn
where ${\mathbf{B}}_r$ is any closed cube of diagonal $r$ centered at the origin in ${\mathbb{R}}^n$  with sides parallel to the coordinate axes, and ${\mathcal{E}}_{{\mathbf{B}}_r } ({\mathbf{x}})$ is the indicator function of ${\mathbf{B}}_r $. 

To construct the space, fix $n$, and let $\mathbb{Q}^n$ be the set $
\left\{ {{\mathbf{x}} = (x_1 ,x_2  \cdots ,x_n ) \in {\mathbb{R}}^n } \right\}$ such that $
x_i$ is rational for each $i$.  Since this is a countable dense set in ${\mathbb{R}}^n $, we can arrange it as $\mathbb{Q}^n  = \left\{ {{\mathbf{x}}_1, {\mathbf{x}}_2, {\mathbf{x}}_3,  \cdots } \right\}$.  For each $l$ and $i$, let ${\mathbf{B}}_l ({\mathbf{x}}_i ) $ be the closed cube centered at ${\mathbf{x}}_i$,  with sides parallel to the coordinate axes and diagonal $r_l  = 2^{ - l}, l \in {\mathbb{N}}$.  Now choose the  natural order which maps $\mathbb{N} \times \mathbb{N}$ bijectively to $\mathbb{N}$:
\[
\{(1,1), \ (2,1), \ (1,2), \ (1,3), \  (2,2), \  (3,1), \ (3,2), \ (2,3), \  \cdots \}.
\]
Let $\left\{ {{\mathbf{B}}_{k} ,\;k \in \mathbb{N}}\right\}$ be the resulting set  of (all) closed cubes $
\{ {\mathbf{B}}_l ({\mathbf{x}}_i )\;\left| {(l,i) \in \mathbb{N} \times \mathbb{N}\} } \right.
$
centered at a point in $\mathbb{Q}^n $, and let ${\mathcal{E}}_k ({\mathbf{x}})$ be the indicator function of ${\mathbf{B}}_k $, so that ${\mathcal{E}}_k ({\mathbf{x}})$ is in ${\mathbf{L}}^p [{\mathbb{R}}^n ] \cap {\mathbf{L}}^\infty  [{\mathbb{R}}^n ] $ for $1 \le p < \infty$.  Define $F_{k} (\; \cdot \;)$ on $
{\mathbf{L}}^1 [{\mathbb{R}}^n ] $ by
 \beqn
F_{k} (f) = \int_{{\mathbb{R}}^n } {{\mathcal{E}}_{k} ({\mathbf{x}})f({\mathbf{x}})d{\mathbf{x}}}. 
\eeqn
It is clear that $F_{k} (\; \cdot \;)$ is a bounded linear functional on ${\mathbf{L}}^p [{\mathbb{R}}^n ] $ for each ${k}$, $\left\| {F_{k} } \right\|_\infty   \le 1$ and, if $F_k (f) = 0$ for all ${k}$, $f = 0$ so that $\left\{ {F_{k} } \right\}$ is fundamental on ${\mathbf{L}}^p [{\mathbb{R}}^n ] $ for $1 \le p \le \infty$ .
Fix ${t_{k}}> 0 $ such that ${\sum\nolimits_{k = 1}^\infty  {t_k}}=1$  and define a measure $d{\mathbf{P}}  ({\mathbf{x}},{\mathbf{y}})$ on ${\mathbb{R}}^n \, \times {\mathbb{R}}^n $ by: 
\[
d{\mathbf{P}}  ({\mathbf{x}},{\mathbf{y}}) = \left[ {\sum\nolimits_{k = 1}^\infty  {t_k {\mathcal{E}}_k ({\mathbf{x}}){\mathcal{E}}_k ({\mathbf{y}})} } \right]d{\mathbf{x}}d{\mathbf{y}}.
\]
 We now define an inner product $\left( {\; \cdot \;} \right) $ on ${\mathbf{L}}^1 [{\mathbb{R}}^n ] $ by
\beqn
\begin{gathered}
 \left( {f,g} \right) = \int_{\mathbb{R}^n  \times \mathbb{R}^n } {f({\mathbf{x}})g({\mathbf{y}})^ *  d{\mathbf{P}}  ({\mathbf{x}},{\mathbf{y}})}  \hfill \\
{\text{             }} = \sum\nolimits_{k = 1}^\infty  {t_k } \left[ {\int_{\mathbb{R}^n } {{\mathcal{E}}_k ({\mathbf{x}})f({\mathbf{x}})d{\mathbf{x}}} } \right]\left[ {\int_{\mathbb{R}^n } {{\mathcal{E}}_k ({\mathbf{y}})g({\mathbf{y}})d{\mathbf{y}}} } \right]^ *.   \hfill \\ 
\end{gathered} 
\eeqn
The choice of $t_k$  is suggested by physical analysis in another context (see Gill and Zachary [GZ]).  We call the completion of ${\mathbf{L}}^1 [{\mathbb{R}}^n ] $, with the above inner product, the Kuelbs-Steadman space, $\mathbf{KS}^2 [{\mathbb{R}}^n ] $.   Steadman [ST] constructed a version of this space by adapting an approach developed by Kuelbs [KB] for other purposes.  \begin{thm} The space $ \mathbf{KS}^2 [{\mathbb{R}}^n ]$ contains  ${\mathbf{L}}^p [{\mathbb{R}}^n ]$ (for each $p,\;1 \leqslant p \leqslant \infty$)  as continuous, compact, dense embeddings.
\end{thm}
\begin{proof}
The proof of the first part is easy, if we notice that ${\mathbf{L}}^1 [{\mathbb{R}}^n ] \cap {\mathbf{L}}^p [{\mathbb{R}}^n ]$ is dense for $1 \le p  < \iy$.  If $f \in {\mathbf{L}}^{\iy} [{\mathbb{R}}^n ]$, then $\left| {\int_{B_k } {f({\mathbf{x}})d{\mathbf{x}}} } \right|^2  \leqslant \left\| f \right\|_{L^\infty  }^2$ for all $k$, so that $\left\| f \right\|_{KS^2 }  \leqslant \left\| f \right\|_{L^\infty  } $.   The proof of compactness follows from the fact that, if $\{f_n \}$ is any weakly convergent sequence in  ${\mathbf{L}}^{p} [{\mathbb{R}}^n ]$ with limit $f$, then  ${\mathcal{E}}_k ({\mathbf{x}}) \in {\mathbf{L}}^{q} [{\mathbb{R}}^n ], \ 1 < q \le \iy$, so that
\[
\int_{\mathbb{R}^n } { {\mathcal{E}}_k ({\mathbf{x}})\left[ {f_n ({\mathbf{x}}) - f({\mathbf{x}})} \right]d{\mathbf{x}}}  \to 0
\]
for each $k$.  Thus,  $\{f_n \}$ converges strongly to $f$ in $ \mathbf{KS}^2[{\mathbb{R}}^n ]$.  Finally, note that $d{\mu}_k ={\mathcal{E}}_k ({\mathbf{x}})d{\bf x}$ defines a measure in $ \mathfrak{M}[{\R}^n ]$, the dual space of ${\mathbf{L}}^\infty  [{\mathbb{R}}^n ]$ and that ${\bf{KS}}^2 [{\R}^n ] \supset {\mathbf{L}}^1 [{\R}^n ]^{ *  * } {\kern 1pt}  = \mathfrak{M}[{\R}^n ] $.
  
\end{proof}
The fact that ${\mathbf{L}}^\infty  [{\mathbb{R}}^n ] \subset {\bf{KS}}^2 [{\mathbb{R}}^n ]$, while $
{\bf{KS}}^2 [{\mathbb{R}}^n ] $ is separable makes it clear in a very forceful manner that separability is not an inherited property. 

It is of particular interest that ${\bf{KS}}^2 [{\R}^n ] \supset \mathfrak{M}[{\R}^n ] $, the space of bounded finitely additive set functions defined on the Borel sets $\mathfrak{B}[{\R}]^n$.
Recall that $\mathfrak{M}[{\mathbf{R}}^n]$ contains the Dirac delta measure and the free-particle Green's function for the Feynman integral.  Thus, ${\bf{KS}}^2 [{\R}^n ]$ contains the Dirac measure and the kernel for the Feynman integral as norm bounded elements (the original reason for our interest).  It is clear from Theorem 1.4  that the convolution operator  has a bounded extension to ${\bf{KS}}^2 [{\R}^n ]$. This result was used in \cite{GZ1} to prove that the path integral could be rigorously constructed in exactly the manner envisioned by Feynman (see also \cite{GZ2}).  
\begin{thm} The Hilbert space ${\bf{KS}}^2 [{\R}^n ]$ satisfies ${\mcB}'  \subset {\mcH}_2$ for the following classical Banach spaces:
\begin{enumerate} 

\item The bounded continuous functions on ${\R}^n, \;  {\mcC}_b[{\R}^n]$.

\item The bounded uniformly continuous functions on ${\R}^n$, ${\mathbf{UBC}}[\mathbb{R}^n ]$ .

\item The space ${\bf L}^p[\mathbb{R}^n ]$, for  $1 \le p \le \iy$.

\end{enumerate} 
\end{thm}  
\begin{rem}  There is quite a lot of flexibility in the choice of the family of positive numbers $\{t_k\},\; \sum\limits_{k = 1}^\infty  {t_k }  = 1$. This is somewhat akin to the standard metric used for $\R^{\infty}$.  Recall that, for any two points $X,\ Y \in \R^{\infty}, \ d(X,Y)=\sum\limits_{n = 1}^\infty{\tfrac{1}{2^n} \frac{\left|X-Y\right|}{1+\left|X-Y\right|}}$.  The family of numbers $\{ \tfrac{1}{2^n} \}$ can be replaced by any other sequence of positive numbers whose sum is one, without affecting the topology.  We have used physical analysis to choose the family $\{t_k\}$, so they are interpreted as probabilities for the occurrence of a particular discrete path.

There is also some ambiguity associated with the order for  ${\mathbb{Q}}_n$ and the order for $\N \times \N$.  (We have used simplicity to choose the order for $\N \times \N$.)   For our work, the important fact is that, for any combination of orders, the properties of  $\mathbf{KS}^2 [{\mathbb{R}}^n ]$ are invariant.  
\end{rem} 
\subsection{Weak Integral}
The purpose of this section is to indicate one other benefit that $\mathbf{KS}^2 [{\mathbb{R}}^n ]$ offers for analysis.  Define the  distributional (or weak) integral on $\R$ by (see Talvila \cite{TA}):     

\begin{Def} Let $F'=DF$ be the weak derivative of $F$.
We define
\[
\mathcal{A}_{\mathbf{c}} (\mathbb{R} ) = \left\{ {f = D F\left| {\; ,\;F \in \mathcal{B}_{\mathbf{c}} (\mathbb{R} )} \right.} \right\},
\]
where
\[
\mathcal{B}_{\mathbf{c}} (\mathbb{R} ) = \left\{ {F \in {\mathbf{C}}(\mathbb{R} )\left| {\;\,\;\mathop {\lim }\limits_{{x} \to  - \infty } F(x) = 0,\;\;\mathop {\lim }\limits_{{{x}} \to \infty } F({{x}})  \in \R } \right.} \right\}.
\] 
If $f \in \mathcal{A}_{\mathbf{c}} (\mathbb{R} )$, we say that $F \in \mathcal{B}_{\mathbf{c}} (\mathbb{R} )$ is the weak integral of $f$  and write
\[
F(x) = (w)\int_{ - \infty }^x {f(y)dy} .
\]
\end{Def}
The following is proved in Talvila \cite{TA}.
\begin{thm}With the Alexiewicz norm $\left\| {\, \cdot \,} \right\|_D $, the space ${\mathcal{A}}_{\bf c}[{\mathbb{R}}]$ has the following properties:
\begin{enumerate}
\item ${\mathcal{A}}_{\bf c}[{\mathbb{R}}]$ is a separable Banach space and a Banach lattice, which contains ${\bf L}^1[{\mathbb{R}}]$ and the Denjoy integrable functions (restricted and wide sense) as dense subsets.
\item ${\mathcal{A}}_{\bf c}[{\mathbb{R}}]$ is isometrically isomorphic to ${\mathcal{B}}_{\bf c}[{\mathbb{R}}]$.
\item ${\mathcal{A}}_{\bf c}[{\mathbb{R}}]$ is the completion of $D(\R)$ (space of Denjoy integrable functions).
\item The dual space ${\mathcal{A}}^*_{\bf c}[{\mathbb{R}}]$ of ${\mathcal{A}}_{\bf c}$, is $\mathcal{BV}(\R)$ (i.e., functions of bounded variation on $\R$).
\end{enumerate}
\end{thm}
This theorem allows us to include  the restricted and wide sense Denjoy integrals in the class of distributions.   
\begin{thm}  The space ${\mathcal{A}}_{\bf c}[{\mathbb{R}}]$ is a continuous dense and  compact embedding in ${\bf{KS}}^2[\R]$. 
\end{thm}
\begin{proof} Since ${\mathcal{E}}_k ({\mathbf{x}}) \in \mathcal{BV}(\R)$ for each $k$, compactness follows.  To prove continuity, note that
\[
\begin{gathered}
  \left| {\int_{B_k } {f(x)dx} } \right|^2  \leqslant \left\| f \right\|_D^2 \;\quad \forall k \Rightarrow  \hfill \\
  \left\| f \right\|_{KS}^2  = \sum\limits_{k = 1}^\infty  {t_k \left| {\int_{B_k } {f(x)dx} } \right|^2 }  \leqslant \sum\limits_{k = 1}^\infty  {t_k \left\| f \right\|_D^2  = \left\| f \right\|_D^2 } . \hfill \\ 
\end{gathered} 
\]
Thus, ${\bf L}^1[{\mathbb{R}}] \subset {\mathcal{A}}_{\bf c}[{\mathbb{R}}] \subset {\bf{KS}}^2[\R]$ is a continuous and dense embedding. 
\end{proof}
\begin{rem}
There is also a weak integral in $\R^n$ (see \cite {ASV} and \cite{MO} for details). If $f \in {\mathcal{D}'}(\R^n)$ then $f$ is integrable if there is a function $F \in \bf{C}(\R^n)$ such that $DF = f$, where $D = \tfrac{{\partial ^n }} {{\partial x_1 \partial x_2  \cdots \partial x_n }}$. Thus, 
\[
\int_{\mathbb{R}^n } {f(x)\varphi (x)} dx = \int_{\mathbb{R}^n } {DF(x)\varphi (x)} dx = ( - 1)^n \int_{\mathbb{R}^n } {F(x)D\varphi (x)} dx,
\]
for all $\phi$ in ${\bf{C}}_c^{\infty}(\R^n)$.
\end{rem}

In this case, we can use our generalization of the  Alexiewicz norm, equation (2.2). Thus,  if $f \in D({\mathbb{R}}^n )$, then
\[
\left\| f \right\|_D  = \sup _{r > 0} \left| {\int_{{\mathbf{B}}_r } {f({\mathbf{x}})d{\mathbf{x}}} } \right| = \sup _{r > 0} \left| {\int_{\mathbf{R}^n } {{\mathcal{E}}_{{\mathbf{B}}_r } ({\mathbf{x}})f({\mathbf{x}})d{\mathbf{x}}} } \right| < \infty, 
\]
to construct the space $ \mathcal{A}_{\mathbf{c}} ({\mathbb{R}}^n )$. 

\subsection{Discussion}
Let ${\bf J}_{KS}(\cdot)$ be the conjugate linear isomorphism between ${\bf KS}^2[\R^n]$ and its dual $\{ {\bf KS}^2[\R^n] \}'$.  Since ${\bf KS}^2[\R^n]$ contains ${\bf L}^p[\R^n], \ 1 \le p  \le \iy$,  for each $f({\bf x}) \in {\bf KS}^2[\R^n], \ {\bf J}_{KS}(f_{\bf x})(\, \cdot \,) = \left\langle {\, \cdot, \ f({\bf x})} \right\rangle$ is a continuous linear functional on all of these spaces.  However, this linear functional need not be in the dual space of any one of them.  Thus, in general, we cannot automatically assume that: 
\[
\mcB \subset \mcH \Rightarrow {\mcH}' \subset {\mcB}'.
\]
\subsection{The Corresponding ${\mcH}_1$}
For completion, in this section we construct the ${\mcH}_1$ version of ${\bf KS}^2$.  

Recall that $\sum _{n = 1}^\infty  \tfrac{1}{{n^2 }} = \tfrac{{\pi ^2 }}{6}$.  Thus,  setting  $\lambda _n  = \tfrac{6}{{\pi ^2 n^2 }}$, we see that  $\sum _{n = 1}^\infty \lambda _n = 1$.  If we let $\{ \varphi _n \}$ be the complete orthonormal set generated by the Hermite functions on $\R^n$, then $\varphi _n  \in \mcB$ for all the classical Banach spaces  in Theorem 1.7.  Thus, we can define ${\bf T}_{12}$ and ${\bf KS}_1^2$ by:
\[
\begin{gathered}
  {\mathbf{T}}_{12} u = \sum\nolimits_{n = 1}^\infty  {\tfrac{6}
{{\pi ^2 n^2 }}\left( {u, \varphi _n } \right)_{{\mathbf{KS}}^2 } \varphi _n } {\text{  and,}} \hfill \\
  {\mathbf{KS}}_1^2  = \left\{ {u \in B\left| {\;\sum\nolimits_{n = 1}^\infty  {\tfrac{{\pi ^2 n^2 }}
{6}\left| {\left( {u, \varphi _n } \right)_{{\mathbf{KS}}^2 } } \right|^2  < \infty } } \right.} \right\},\quad {\text{with}} \hfill \\
  \quad \quad \quad \quad \left( {u,v} \right)_{{\mathbf{KS}}_1^2 }  = \sum\nolimits_{n = 1}^\infty  {\tfrac{{\pi ^2 n^2 }}
{6}\left( {u, \varphi _n } \right)_{{\mathbf{KS}}^2 } \left( {\varphi _n ,v } \right)_{{\mathbf{KS}}^2 } .}  \hfill \\ 
\end{gathered}
\]
We call ${\mathbf{KS}}_1^2$ the Gross-Steadman space.  Historically, Gross \cite{GR} first proved that every real separable Banach space contains a separable Hilbert space as a dense embedding, and that this space is the support of a Gaussian measure.  This was a major extension of Wiener's theory, which  was based on the use of the (densely embedded Hilbert)  Sobolev space ${\bf H}^{1}[0,1]\subseteq {\bf C}[0,1]$ (i.e., $u \in {\bf H}^{1}[0,1]$ means that its first order weak derivative is in ${\bf C}[0,1]$). Motivated by Gross' theorem, Kuelbs realized that the inclusion ${\bf H}^{1}[0,1]\subseteq {\bf C}[0,1] \subset {\bf L}^2[0,1]$ might have an extension and prove the original version of Theorem 1.2. 

While ${\mathbf{KS}}^2 \ (={\mcH}_2)$ will be explicitly used during the remainder of the paper, ${\mathbf{KS}}_1^2 \ (={\mcH}_1)$ will be equally implicit, and $\mcB$ shall always refer to one of the Banach spaces in Theorem 2.3.
\section{\bf Closed Operators}
\begin{Def} A Banach space $\mcB$ is said to be:
\begin{enumerate}
\item quasi-reflexive if $ \dim \left\{ {{{{\mcB}''} \mathord{\left/ {\vphantom {{{\mcB}''} {\mcB}}} \right.
 \kern-\nulldelimiterspace} {\mcB}}} \right\} < \infty $, and 
 \item nonquasi-reflexive if $ \dim \left\{ {{{{\mcB}''} \mathord{\left/ {\vphantom {{{\mcB}''} {\mcB}}} \right.
 \kern-\nulldelimiterspace} {\mcB}}} \right\} = \infty $.
 \end{enumerate}
\end{Def}
In general,  it is not reasonable to expect that Theorem 1.4 will hold for all operators in $\mcC[\mcB]$.  An important result by Vinokurov, Petunin and Pliczko \cite{VPP} shows that for every nonquasi-reflexive Banach space $\mcB$
 (for example, $C[0; 1]$ or $L^1[{\R}^n], \; n \in \N$), there is a bounded linear injective operator $A^{-1}$ with a dense range whose inverse $A$ is a closed densely defined linear operator which is not of the first Baire class. This means that there does not exist a sequence of bounded linear operators $A_{n} \in L[\mcB]$ such that, for $x \in D(A), \; A_n x \rightarrow Ax$, as $n \rightarrow \iy$.

Recall that a m-dissipative linear operator is the generator of a $C_0$-contraction semigroup and $Ran(\la I-A)=\mathcal{B}$ for every $\la >0$ (see Pazy \cite{PZ}).  Furthermore, the Yosida approximator \cite{YS}, $A_{\la}= \la AR(\la , A)$, is a bounded linear operator which converges strongly to $A$ on $D(A)$.

\begin{thm} Let $A \in  \mcC[\mcB]$, with ${\mcB}' \subset {\mcH}_2$.  The operator $A$ is in the first Baire class if and only if it has an adjoint $A^{*}$.
\end{thm}
\begin{proof} Let ${\mcH}_1 \subset {\mcB} \subset {\mcH}_2$ as in Theorem 1.2, and suppose that $A$ has an adjoint $A^* \in \mcC[\mcB]$. Let $T= [A^{*}A]^{1/2}, \; {\bar T}=[AA^{*}]^{1/2}$ (the negatives of each generate $C_0$-contraction semigroups).   Since $T$ is nonnegative, it follows that ${I+\al T}$ has a bounded inverse $S(\al)=(I+\al T)^{-1}$, for $\al >0$.  It is also easy to see that  $AS(\al)$ is bounded and, on $D(A), \ AS(\al)={\bar S}(\al )A=(I+\al {\bar T})^{-1}A$ (see Kato \cite{K}, pages 335 and 481). Using this result, we have:
\[
\mathop {\lim }\limits_{\alpha  \to 0^ +  } AS(\alpha )x = \mathop {\lim }\limits_{\alpha  \to 0^ +  } {\bar S}(\alpha )Ax= Ax, \; {\rm for} \; x \in D(A).
\]
It follows that $A$ is in the first Baire class.

To prove the converse, suppose that $A$ is in the first Baire class.  Thus,  there is a sequence of bounded linear operators $\{A_n \}$ such that, for $x \in D(A), \ A_n x \rightarrow Ax$ as $n \rightarrow \iy$ .  Since each $A_n$ is bounded, by Theorem 1.3, each $A_n$ has an adjoint $A^*_n$ and both can be extended to bounded linear operators ${\bar A}_n, \ {\bar A}_n^*$ on ${\mathcal{H}}_2$ (by Theorem 1.4).  Furthermore, we have 
$\lt\| {\bar A}_n \rt\|_{{\mcH}_2} \le k \lt\|A_n \rt\|_{\mcB}$ 
and $\lt\|{\bar A}_n^* \rt\|_{{\mcH}_2} \le k\lt\|A_n^{*} \rt\|_{\mcB}$.  It follows that the sequence $\{ {\bar A}_n x \}$ converges for each $x \in D(A)$.   If we define ${\bar A}$  as the closure in ${\mcH}_2$ of  $lim_{n \rightarrow \iy} {\bar A}_n x$ for $x \in D(A)$, then  ${\bar A} \in \mcC[{\mcH}_2]$.  

Since ${\bar A}$ is a closed densely defined linear operator, its ${\mcH}_2$ adjoint, ${\bar A}^*$ is densely defined and ${\bar A}= {\bar A}^{**}$ (see Rudin \cite{RU}, Theorem 13.12, page 335).  From this, we see that  ${\bar A}^{*}$ is a closed densely defined linear operator on ${\mcH}_2$.  Since  ${\bar A}$ restricted to $\mcB$ is $A, \ {\bar A}^{*}$ restricted to $\mcB$ defines $A^*$.
\end{proof}
If $\mcB$ is a quasi-reflexive separable Banach space, it is shown in \cite{VPP} that every bounded linear injective operator $A^{-1}$ with a dense range whose inverse $A$ is a closed densely defined linear operator is of the first Baire class.  Since, to our knowledge, every operator $A \in {\mcC}[{\mcB}]$ cannot be obtained from an  $A^{-1}$ in the class of bounded linear injective operators with a dense range, it's still not known if all operators in ${\mcC}[{\mcB}]$ are of the first Baire class (even if $\mcB$ is reflexive). Thus, although the theorems we prove in this section hold for all operators of first Baire class, we restrict our consideration to  generators of $C_0$-contraction semigroups.  
\begin{thm} If $A$ generates a $C_0$-contraction semigroup and ${\mathcal{B}}'  \subset \mathcal{H}_{2}$, then:
\begin{enumerate}
 \item $A$ has a closed densely defined extension $\bar{A}$ to ${\mcH}_2$, which is also the generator of a $C_0$-contraction semigroup.  
\item $\rho (\bar A) = \rho (A)$ and $\sigma (\bar A) = \sigma (A)$.
\item The adjoint of $\bar{A}, \, {\bar{A}}^*$, restricted to $\mcB$, is the adjoint $A^*$ of $A$, that is:
\begin{itemize}
\item[-] the operator $A^ * A \geqslant 0$,	
\item[-]  $(A^ *  A)^ *   = A^ *  A$ and 
\item[-] $I + A^ * A$ has a bounded inverse.
 \end{itemize} 
\end{enumerate} 
\end{thm}
\begin{proof}

{\bf{Part I}}
\newline
Let $T(t)$ be the semigroup generated by $A$.  By Theorem 1.4,  as a bounded linear operator, $T(t)$ has a bounded extension $\bar{T}(t)$ to ${\mathcal{H}}_2$.  

We prove that $\bar{T}(t)$ is a $C_0$-semigroup.  (The fact that it is a contraction semigroup will follow later.)  It is clear that $\bar{T}(t)$ has the semigroup property.  To prove that it is strongly continuous, use the fact that $\mcB$ is dense in ${\mcH}_2$ so that, for each $u \in {\mcH}_2$, there is a sequence  $\{u_n \}$  in $\mcB$ converging to $u $.  We then have: 
\[
\begin{gathered}
  \mathop {\lim }\limits_{t \to 0} \left\| {\bar T(t) u - u} \right\|_2  \leqslant \mathop {\lim }\limits_{t \to 0} \left\{ {\left\| {\bar T(t)u - \bar T(t)u_n } \right\|_2  + \left\| {\bar T(t) u_n  - u_n } \right\|_2 } \right\} + \left\| {u_n  - u} \right\|_2  \hfill \\
   \leqslant k \left\| {u - u_n } \right\|_2  + \mathop {\lim }\limits_{t \to 0} \left\| {\bar T(t) u_n  - u_n } \right\|_2  + \left\| {u_n  - u} \right\|_2  \hfill \\
   = (k + 1)\left\| {u - u_n } \right\|_2  + \mathop {\lim }\limits_{t \to 0} \left\| {T(t) u_n  - u_n } \right\|_2  = (k + 1)\left\| {u - u_n } \right\|_2 , \hfill \\ 
\end{gathered} 
\]
where we have used the fact that $\bar{T}(t) u_n=T(t) u_n$ for $u_n \in \mcB$, and $k$ is the constant in Theorem 1.4.  It is clear that we can make the last term on the right as small as we like by choosing $n$ large enough, so that $\bar{T}(t)$ is a $C_0$-semigroup.
  
To prove (1), note that, if $\bar A$ is the extension of $A$, and $\lambda I - \bar A$ has an inverse, then $\lambda I - A$ also has one, so $\rho (\bar A) \subset \rho (A)$
 and $Ran(\lambda I - A)_{\mcB} \subset Ran(\lambda I - \bar A)_{{\mcH}_2} \subset \overline {Ran(\lambda I - A)}_{{\mcH}_2}$ for any $\lambda  \in {\mathbb{C}}$.  For the other direction, note that, since $A$ generates a $C_0$-contraction semigroup, $\rho (A) \ne \emptyset$.  Thus, if $\lambda  \in \rho (A)$, then $(\lambda I - A)^{ - 1} $ is a continuous mapping from $Ran(\lambda I - A)$ onto $D(A)$ and $Ran(\lambda I - A)$ is dense in ${\mathcal{B}}$.  Let $u  \in D(\bar A) $, so that $(u ,{\bar{A}} u ) \in \hat G(A)$, the closure of the graph of $A$ in ${\mcH}_2$.  Thus, there exists a sequence $\{ u _n \}  \subset D(A) $ such that
$\left\| {u  - u_n } \right\|_G  = \left\| {u  - u _n } \right\|_{{\mcH}_2}  + \left\| {\bar A u  - \bar A u _n } \right\|_{{\mcH}_2}  \to 0
$ as $n \to \infty $.  Since $\bar A u _n= A u _n$, it follows that $(\lambda I - \bar A) u  = \lim _{n \to \infty } (\lambda I - A) u _n $.  However, by the boundedness of $(\lambda I - A)^{ - 1} $
 on $R(\lambda I - A)$ we have that, for some $\delta  > 0$,
\[
\left\| {(\lambda I - \bar A) u } \right\|_{{\mcH}_2}  = \lim _{n \to \infty } \left\| {(\lambda I - A) u_n } \right\|_{{\mcH}_2}  \ge \lim _{n \to \infty } \delta \left\| {u _n } \right\|_{{\mcH}_2}  = \delta \left\| u  \right\|_{{\mcH}_2}.
\]
It follows that $\lambda I - \bar A$ has a bounded inverse and, since $D(A) \subset D(\bar A)$ implies that $Ran(\lambda I - A) \subset Ran(\lambda I - \bar A)$, we see that $Ran(\lambda I - \bar A)$ is dense in ${\mathcal{H}}_2$ so that $\lambda  \in \rho (\bar A)$  and hence $\rho (A) \subset \rho (\bar A)$.  It follows that $\rho (A) = \rho (\bar A)$ and necessarily, $\sigma (A) = \sigma (\bar A)$.
 
Since $A$ generates a $C_0$-contraction semigroup, it is m-dissipative.  From the Lumer-Phillips Theorem (see Pazy [PZ]), we have  that $Ran(\lambda I - A)=\mcB$ for $\la >0$.  It follows that $\bar A$ is m-dissipative and $Ran(\lambda I - \bar A)= {{\mcH}_2}$. Thus, $\bar{T}(t)$ is a $C_0$-contraction semigroup.

We now observe that the same proof applies to ${\bar{T}}^*(t)$, so  that ${\bar{A}}^*$ is also the generator of a $C_0$-contraction semigroup on ${{\mcH}_2}$.  

Clearly  ${\bar{A}}^*$ is the adjoint of ${\bar{A}}$ so that, from 
von Neumann's Theorem, ${\bar{A}}^*{\bar{A}}$ has the expected properties.  By a result of Kato \cite{K} (see page 276), 
$\bar{\bf D}=D({\bar{A}}^*{\bar{A}})$ is a core for  ${\bar{A}}$ (i.e., the set of elements $\{u, \, {\bar{A}} u \}$ is dense in the graph, $G[{\bar{A}}]$, of ${\bar{A}}$ for $u \in \bar{\bf  D}$).  From here, we see that the restriction $A^*$ of ${\bar{A}}^*$ to $\mcB$ is the generator of a $C_0$-contraction semigroup and ${\bf D}=D({{A}}^*{{A}})$ is a core for  ${{A}}$.    The proof of (3) for ${{A}}^*{{A}}$  now follows.  
\end{proof}

\begin{thm} Let  $A \in {\mcC}[{\mcB}]$ be the generator of a $C_0$-contraction semigroup.  If ${\mcB}' \subset {\mcH}_2$, then there exist a m-accretive operator $R$ and a partial isometry $W$ such that $A=WR$ and $D(A)=D(R)$.
\end{thm}
\begin{proof}
The fact that ${\mcB}' \subset {\mcH}_2$ ensures that $A^*A$ is a closed selfadjoint operator on ${\mcB}$ by Theorem 3.3.  Furthermore, both $A$ and $A^*$ have closed densely defined extensions  $\bar A$ and ${\bar A}^*$ to ${\mathcal H_2}$. Thus, the operator $\hat R=[{\bar A}^*{{\bar A}}]^{1/2}$ is a well-defined m-accretive selfadjoint linear operator on ${\mathcal H_2},\, {\bar A}={\bar W}{\bar R}$ for some partial isometry ${\bar W}$ defined on ${\mathcal H_2}$, and $D({\bar A})=D({\bar R})$. Our proof is complete when we notice that the restriction of ${\bar A}$  to ${\mathcal B}$ is ${A}$ and ${\bar R}^2$ restricted to ${\mathcal B}$ is ${A^*A}$, so that the restriction of ${\bar W}$ to ${\mathcal B}$ is well-defined and must be a partial isometry.  The equality of the domains is obvious.
\end{proof}
\subsection{Operators on $\mcB$} 
\begin{Def} Let $S$ be bounded, let $A$ be closed and densely defined, and let ${\mathcal{U}},\,{\mathcal{V}}$ be subspaces of ${\mathcal{B}}$:
\begin{enumerate}
\item $A$ is said to be naturally self-adjoint if $A = A^*$ on $D(A)$.
\item  $A$ is said to be normal if $AA^*   = A^*  A$ on $D(A)$.
\item $S$ is unitary if $SS^*   = S^* S = I$.
\item The subspace ${\mathcal{U}}$ is $ \bot $ to ${\mathcal{V}}$ if for each $v \in {\mathcal{V}}$ and $\forall u  \in {\mathcal{U}},\; \left\langle {v , f_u ^s } \right\rangle  = 0$ and, for each $u \in {\mathcal{U}}$ and $\forall v  \in {\mathcal{V}}, \; \left\langle {u , f_v ^s } \right\rangle  = 0$.
\end{enumerate}
\end{Def}
The last definition is transparent since, for example,
\[
\left\langle {v ,f_u ^s } \right\rangle  = 0 \Leftrightarrow \left\langle {v ,J_2 (u)} \right\rangle  = \left( {v ,u } \right)_2  = 0\;\;\forall v  \in \mathcal{V}.
\]
With respect to our definition of natural selfadjointness,
the following related definition is due to Palmer \cite{PL}, where the operator is called symmetric.  This is essentially the same as a Hermitian operator as defined by  Lumer \cite{LU}.
\begin{Def}
A closed densely defined linear operator $A$ on ${\mathcal{B}}$ is called self-conjugate if both $iA$ and $-iA$ are dissipative.
\end{Def}
\begin{thm}(Vidav-Palmer)  A linear operator $A$, defined on ${\mathcal{B}}$, is self-conjugate if and only if $iA$ and $-iA$ are generators of isometric semigroups.
\end{thm}
\begin{thm}
The operator  $A$, defined on $\mathcal B$, is self-conjugate if and only if it is naturally self-adjoint.
\end{thm}
\begin{proof} Let $\bar A$ and ${\bar A}^*$ be the closed densely defined extensions of $A$ and $A^*$ to ${\mathcal H_2}$.  On  ${\mathcal{H}}_2$, $\bar A$ is naturally self-adjoint if and only if $i \bar A$ generates a unitary group, if and only if it is self-conjugate.  Thus, both definitions coincide on ${\mathcal{H}}_2$.  It follows that the restrictions coincide on ${\mathcal{B}}$.
\end{proof}
For later reference, we note that orthogonal subspaces in $
{\mathcal{H}}_2$ induce orthogonal subspaces in ${\mathcal{B}}$.
\begin{thm}   (Gram-Schmidt) If ${\mathcal{B}}$ has a basis $\{ \varphi _i ,\,\,1 \leqslant i < \infty \} $ then there is an orthonormal basis $\{ \psi _i ,\,\,1 \leqslant i < \infty \} $ for ${\mathcal{B}}$ with a corresponding set of orthonormal duality maps $\{ f_{\psi _i}^s ,\,\,1 \leqslant i < \infty \} $
 (i.e., $\left\langle {\psi _i ,f_{\psi _i}^s } \right\rangle  = \delta _{ij}$). 
\end{thm}
\begin{proof}Since each $\varphi _i $ is in ${\mathcal{H}}_2$, we can construct an orthogonal set of vectors $\{ \phi _i ,\,\,1 \leqslant i < \infty \} $ in ${\mathcal{H}}_2$ by the standard Gram-Schmidt process.  Set $\psi _i  = {{\phi _i } \mathord{\left/
 {\vphantom {{\phi _i } {\left\| {\phi _i } \right\|}}} \right.
 \kern-\nulldelimiterspace} {\left\| {\phi _i } \right\|}}_\mathcal{B} $
 and ${\hat f}_{\psi _i}^s  = {{J (\psi _i )} \mathord{\left/
 {\vphantom {{J (\psi _i )} {\left\| {\psi _i } \right\|}}} \right.
 \kern-\nulldelimiterspace} {\left\| {\psi _i } \right\|}}_\mathcal{H}^2 $ on the subspace $M=< {\psi _i}>$.  Now use the Hahn-Banach Theorem to extend ${\hat f}_{\psi _i}^s$ to all of $\mcB$ as in Section 1, to get ${f}_{\psi _i}^s$. 
From here, it is easy to check that $\{ \psi _i ,\,\,1 \leqslant i < \infty \} $ is an orthonormal basis for ${\mathcal{B}}$ with corresponding orthonormal duality maps $\{ {f}_{\psi _i}^s ,\,\,1 \leqslant i < \infty \} $.  
\end{proof}
We close this section with the following observation about the use of ${\bf KS}^2$.  Let $A$ be any closed densely defined positive linear operator on $\mcB$ with a discrete positive spectrum $\{ \la_i \}$. In this case, $-A$ generates a $C_0$-contraction semigroup, so that it can be extended to ${\mcH}_2$ with the same properties. If we compute the ratio $\tfrac{{\left\langle {A\psi ,f_\psi ^s } \right\rangle }}
{{\left\langle {\psi ,f_\psi ^s } \right\rangle }} $ in $\mcB$, it will be ``close" to the value of 
$\tfrac{{\left( {{\bar A} \psi ,\psi } \right)_{{\mcH}_2 } }}
{{\left( {\psi ,\psi } \right)_{{\mcH}_2 } }}$ in ${{\mcH}_2 }$.  On the other hand, note that we can use the min-max theorem on $\mcH_2$ to compute the eigenvalues and eigenfunctions of $A$ via $\bar A$ exactly on $\mcH_2$.  Thus, in this sense, the min-max theorem holds on $\mcB$.
\section{\bf Extension of the Poincar\'{e} inequality} 
\subsection{Introduction}
There are a number of versions of the Poincar\'{e} inequality (see Evans \cite{EV}).  We consider the version that naturally appears in the theory of Markov processes.  Let $\mu$ be a Borel probability measure associated with the transition semigroup $S(t)$ for a given Markov process with generator $A$.   The measure $\mu$ is called an invariant measure if:
\[
\int_{\mathbb{R}^3 } {S(t)u({\mathbf{x}})d\mu ({\mathbf{x}})}  = \int_{\mathbb{R}^3 } {u({\mathbf{x}})d\mu ({\mathbf{x}})} ,\quad t > 0,
\]
for any $u({\bf x}) \in {\mathbb{C}_{c}^{\infty}}[\R^3]$.  If $u$ is any function in $L_{}^{p}[\R^3, d{\mu}]$ and we set
$\bar u=\int_{\mathbb{R}^3 } {u({\mathbf{x}})d\mu ({\mathbf{x}})}$, it is known that  for $1 \le p < \infty$:
\[
\lim _{t \to \infty } \left\| {S(t)u - \bar u} \right\|_p  = 0.
\]
Since the generator of $S(t)$ is strongly elliptic, if $u \in W_{\mu}^{1,p}[\R^3, d{\mu}]$ (the space of functions whose first order weak derivative is in $L_{}^{p}[\R^3, d{\mu}]$), the Poincar\'{e} inequality states that:
\beqn
\int_{\mathbb{R}^3 } {\left| {u - \bar u} \right|^p d\mu }  \le C\int_{\mathbb{R}^3 } {\left| {Du}({\bf{x}}) \right|^p d{\mu}({\bf{x}})},
\eeqn
where $C$ is a positive constant and $\bar{u} =\int_{\R^3}u({\bf{x}}) \, d{\mu}({\bf{x}})$.    
\subsection{Purpose}
The purpose of this section is to show that our adjoint theory allows us to  extend equation (4.1) to a large class of operators, which includes all $C_0$-generators $A= -WR, \; R=-[A^*A]^{1/2}$, where the spectrum of $R$ is bounded away from  zero.  

In this section, we assume that  $\bar u=0$, so that  $\left\| u \right\|_p^p  \le C\left\| {Du} \right\|_p^p$.
\subsection{Hilbert space case}
We first assume that we are working on a separable Hilbert space $\mcH$. In this case, for any closed densely defined linear operator $A$, both $R=- [A^{*}A]^{1/2}$ and ${\bar R}= - [AA^{*}]^{1/2}$ are generators of $C_0$-analytic contraction semigroups on ${\mathcal{H}}$.  Furthermore, there is a unique partial isometry $W$  such that $A=-WR=-{\bar R}W$, and  $A^*=-W^*{\bar R} =-RW^* $, see Kato \cite{K}, page 334.  (It should be noted that $A$ itself is rarely  a $C_0$-semigroup generator of any type.)

\begin{thm} Let $S(t)$ be the analytic contraction semigroup generated by $R$.  If, for $u \in \mathcal{H}$, there is a $T\in (0,\iy)$ such that, for $t \ge T$,
$
\left\| {S(t)u} \right\|_{\mathcal{H}}  \le 
r\left\| u \right\|_{\mathcal{H}}
$,  with $r<1$, then there exists a constant $c$ such that, for each $u \in D(A)$, 
$\left\| u \right\|_\mathcal{H}  
\le c \left\| {Au} \right\|_{\mathcal{H}}$.
\end{thm}
\begin{proof} Since $S(t)$ is analytic, with $R$ as its generator, we have $RS(t)u= S(t)Ru$ for $u \in D(A)$. Thus,
\[
\begin{gathered}
  \left| {\left\| {S(t)u} \right\|_\mathcal{H}  - \left\| u \right\|_\mathcal{H} } \right| \le \left\| {S(t)u - u} \right\|_\mathcal{H}  = \left\| {\int_0^t {RS(\tau )u d\tau } } \right\|_\mathcal{H}   \hfill \\
 = \left\| {\int_0^t {S(\tau )Ru d\tau } } \right\|_\mathcal{H}   \le  \int_0^t {\left\| {S(\tau )Ru} \right\|_\mathcal{H} d\tau }  \le t\left\| {Ru} \right\|_\mathcal{H} =t\left\| {Au} \right\|_\mathcal{H}.  \hfill \\ 
\end{gathered} 
\]
Hence, for
$
t \ge T
$,
\[
\left\| u \right\|_\mathcal{H}  - r\left\| u \right\|_\mathcal{H} \le  \left\| u \right\|_\mathcal{H}  - \left\| {S(t)u} \right\|_\mathcal{H}  \le T\left\| {Au} \right\|_{\mathcal{H}}.
\]
If we set
$
c = \tfrac{T}{{1 - r}},
$
then $\left\| u \right\|_\mathcal{H}  
\le c \left\| {Au} \right\|_{\mathcal{H}}$.
\end{proof}
(Note that the proof of Theorem 4.1 does not depend on the Hilbert space structure.)

The natural question is: What are the additional conditions on $R$ that make the above result possible?
The following conditions (for separable Banach spaces) are known (see Pazy \cite{PZ}):
\begin{thm}  Let $\mcB$ be a separable Banach space. If:
\begin{enumerate}
\item for some $p, \ 1 \le p  < \iy$
\[
\int_0^\infty  {\left\| {S(t)u} \right\|_{\mcB}^p dt}  < \infty \quad {\text{for every }}u \in B,\;\;{\text{or}}
\]
\item $S(t)$ is an analytic contraction semigroup whose generator $R$ has a spectrum  $\sigma (R)$, such that
\beqn
\sigma  = \sup \left\{ {\operatorname{Re} (\lambda ):\;\lambda  \in \sigma (R)} \right\} < 0,
\eeqn
\end{enumerate}
then there  are constants $M \ge 1$ and $ \mu>0 $ such that 
\beqa
\left\| {S(t)} \right\|_{\mathcal{B}} \le Me^{-\mu t}.
\eeqa
\end{thm}
Slemrod \cite{SL}, has proved a general result assuring that $\left\| {S(t)} \right\|_{\mathcal{B}} \le Me^{-\mu t}$.  The following applies to our case.
\begin{thm}  Let $S(t)$ be a semigroup on $\mathcal{H}$.  If either condition of Theorem 4.2 holds,  then there  exists a constant $r, \ 0 < r < 1 $, such that 
\beqn
 \left\| {S(t)} \right\|_{\mathcal{H}} \le r.
\eeqn
\end{thm}
\begin{proof} Under the stated conditions, $\left\| {S(t)} \right\|_{\mathcal{H}} \le Me^{-\mu t}$.  If we choose $T> \tfrac{\ln M}{\mu}$  and $r=Me^{-\mu T}$, it is easy to check that inequality (4.3) is satisfied.
\end{proof}
The above theorem applies to all closed densely defined linear operators $A$ such that $A^*A$ is a strictly positive operator, where $R=-[A^*A]^{1/2}$.  In this case, if we drop the analytic condition, the theorem does not hold (see Pazy \cite{PZ}, example 4.2, page 117).
\subsection{Banach space case}
In case we have a separable Banach space $\mcB$, we assume that $A$ is the generator of a $C_0$-contraction semigroup and ${\mcB}' \subset {\mcH}_2$.
\begin{thm} Let $A=WR$ and let  $S(t)$ be the analytic contraction semigroup generated by $R$ on $\mcB$.  If, for $u \in \mathcal{B}$, there is a $0<T< \iy$ such that, for $t \ge T$,
$
\left\| {S(t)u} \right\|_{\mathcal{B}}  \le 
r\left\| u \right\|_{\mathcal{B}}
$,  with $r<1$, then there exists a constant $c$ such that, for each $u \in D(A)$, 
$\left\| u \right\|_\mathcal{B}  
\le c \left\| {Au} \right\|_{\mathcal{B}}$.
\end{thm}
\begin{proof}
The proof is the same as for Theorem 4.1.
\end{proof}
\begin{Def} Let $A$ generate a $C_0$-contraction semigroup and let  $B$ be a closed densely defined linear operator on $\mcB$.  We say that $B$ is relatively bounded with respect to $A$ if $D(A) \subset D(B)$ and there are positive numbers $a, \, b$ such that:
\[
\left\| {Bu} \right\|_{\mathcal{B}} \leqslant a\left\| u \right\|_{\mathcal{B}} + b\left\| {Au} \right\|_{\mathcal{B}} \quad {\text{for }}u \in D(A).
\]
\end{Def}
The proof of the next result follows from Theorem 4.4.
\begin{cor} If $B$ is relatively bounded with respect to $A(=WR)$ and zero is bounded away from $\s(R)$, then there is a constant $c$ such that 
\[
\left\| {Bu} \right\|_{\mathcal{B}} \leqslant  c\left\| {Au} \right\|_{\mathcal{B}} \quad {\text{for }}u \in D(A).
\]
\end{cor}
\section{\bf Extension Of The Spectral Theorem}
\subsection{Introduction}
For any selfadjoint operator in ${\mcC}[{\mcH}]$, the following theorem is well-known.  A proof can be found in \cite{DS}, page 1192-99 (see also Reed and Simon \cite{RS} page 263).
\begin{thm} Let $A \in {\mcC}[{\mcH}]$ be a selfadjoint operator,  with spectrum $\s(A) \subset \R$, then there exists a unique regular countably additive  projection-valued ($=$ spectral)  measure ${\bf{E}}(\Om)$ mapping the Borel sets, ${\mathfrak{B}}[\R]$, over $\R$ into $\mcH$ such that, for each $x \in D(A)$, we have:
\begin{enumerate}
\item $D(A)$ also satisfies
\[
D(A) = \left\{ {\left. {x \in \mathcal{H}} \ \right|\;\int_{\sigma (A)} {\lambda ^2 \left( {{\bf{E}}(d\lambda )x,x} \right)_\mathcal{H}  < \infty } } \right\}
\]
and
\item 
\[
Ax = \mathop {\lim }\limits_{n \to \infty } \int_{ - n}^n {\lambda {\bf{E}}(d\lambda )x}, \; {\rm for } \; x \in D(A). 
\]
\item If $g(\cdot)$ is a complex-valued Borel function defined (a.e) on $\R$, then $g(A)\in {\mcC}[{\mcH}]$ and, for  $ x \in D(g(A))= D_g(A)$,  
\[
g(A)x= \mathop {\lim }\limits_{n \to \infty } \int_{ - n}^n {g(\lambda) {\bf{E}}(d\lambda )x},
\]
 where
\[
D_g(A) = \left\{ {\left. {x \in \mathcal{H}} \ \right|\;\int_{\sigma (A)} {\lt|g(\lambda)\rt|^2 \left( {{\bf{E}}(d\lambda )x,x} \right)_\mathcal{H}  < \infty } } \right\}
\]
and $g(A^*)= {\bar g}(A)$.
\end{enumerate}
\end{thm}
It is an exercise to show that ${\bf{E}}(\Om)x$ is of bounded variation.  (For $\Om=(- \iy, \la], \ {\bf{E}}(\la)x$ is called a spectral function and $\{{\bf{E}}(\la) \}$ is called a spectral family.)  

Theorem 5.1 initiated the general study of operators that have a spectral representation (or functional calculus).  This research has moved in many directions.  The Rellich-Titchmarsh-Kato line is concerned with applications to problems in physics and applied mathematics. In this direction, one is interested in concrete detailed information about the spectrum of various specific operators subject to different constraints (see Rellich \cite{RL}, Titchmarsh \cite{TI} and Kato \cite{K}).  Another line of study follows more closely the approach developed by Stone and von Neumann (independently extending the bounded case by HIlbert).  In this direction one seeks to extend Theorem 5.1 to a larger class of operators via operator theory and functional analysis (see Dunford and Schwartz \cite{DS} and Yosida \cite{YS}).  The notes starting on page 2089 (in \cite{DS}) are especially helpful in understanding the history (and the many other approaches).   
\subsection{Background}
Dunford and Schwartz define a spectral operator as one that has a spectral family similar to that defined in Theorem 5.1 for selfadjoint operators. (A spectral operator is an operator with countably additive spectral measure on the Borel sets of the complex plane.) Strauss and Trunk \cite{STT} define a bounded linear operator $A$, on a Hilbert space $\mcH$, to be spectralizable if there exists a non-constant polynomial $p$ such that the operator $p(A)$ is a scalar spectral operator (has a representation as in Theorem 5.1 (2)).  Another interesting line of attack is represented in the book of Colojoar$\breve{a}$  and Foia\c{s} \cite{CF}. where they study the class of generalized spectral operators. Here, one is not opposed to allowing the spectral resolution to exist in a generalized sense, so as to include operators with spectral singularities.  

The following theorem was proven by Helffer and Sj{\"o}strand \cite{HSJ} (see Proposition 7.2):
\begin{thm} Let $g \in {\mcC}_0^{\iy} [\R]$ and let $\hat{g} \in {\mcC}_0^{\iy} [{\bf C}]$ be an extension of $g$, with $\tfrac{{\partial {\hat {g}}}}{{\partial {\hat {z}}}} = 0$ on $\R$.  If $A$ is a selfadjoint operator on $\mcH$, then
\[
g(A) =  - \frac{1}
{\pi }\iint\limits_{\mathbf{C}} {\frac{{\partial \hat{g}}}
{{\partial \bar z}}\left( {z - A} \right)^{ - 1} dxdy}.
\]
\end{thm}
This defines a functional calculus.  Davies \cite{DA} showed that the above formula can be used to define a functional calculus on Banach spaces for a  closed densely defined linear operator $A$, provided $\rho(A) \cap \R=\emptyset$.  In this program the objective is to construct a functional calculus pre-supposing that the operator of concern has a reasonable resolvent. 
\subsection{Problem}
The basic problem that causes additional difficulty is the fact that many bounded linear operators (on ${\mcH}_2$) are of the form
$
A=B+N,
$
where $B$ is normal and $N$ is nilpotent (i.e., there is a $k \in \N$, such that $N^{k+1}=0, \ N^k \ne 0$).  in this case, $A$ does not have a representation with a standard spectral measure.  On the other hand, $T= [{N^*N}]^{1/2}$ is a selfadjoint operator and, there is a unique partial isometry $W$ such that $N=WT$.  If ${\bf E}(\, \cdot \,)$ is the spectral measure associated with $T$, then $W{\bf E}(\Om)x$ is not a spectral measure but, it is a measure of bounded variation.  Thus, we just might be able to find a easier solution to the problem if we willing drop our requirement that the spectral representation be with respect to a spectral measure in the normal sense.

We begin by noting that, in either of the Strauss and Trunk \cite{STT},  Helffer and Sj{\"o}strand \cite{HSJ} or Davies \cite{DA} cases, the operator $A$ is in the first Baire class.  Thus, Theorem 3.2 shows that $A$ has an adjoint and Theorem 3.4 shows that $A=WR$, where $W$ is a partial isometry and $R$ is a nonnegative selfadjoint linear operator.  Before presenting our solution for the Hilbert space case, we need a few results about vector-valued functions of bounded variation.

Recall that a vector-valued function ${\bf{e}}(\la)$ defined on a subset of $\R$ to $\mcH$  is of bounded variation if 
\[
V({\mathbf{e}},\R)=\mathop {sup}\limits_P \left\| {\sum\limits_{i = 1}^n {[{\mathbf{e}}(b_i ) - {\mathbf{e}}(a_i )]} } \right\|,
\] 
where the supremum is  over all partitions $P$ of non-overlapping intervals $(a_i,b_i)$ in $\R$ (see Hille and Phillips \cite{HP} or Diestel and Uhl \cite{DU}).

The next result is proved in Hille and Phillips \cite{HP} (see page 63).
\begin{thm} Let ${{\bf a}}(\la)$ be a vector-valued function from $\R$ to $\mcH$ of bounded variation.  If $h(\la)$ is a continuous complex-valued function on $(a,b) \subset \R$,  then the following holds:
\begin{enumerate}
\item The integral $\int_a^b {h(\lambda )} d{\bf{a}}(\lambda )$ exists in the  $\mcH$ norm.

\item  If $T$ is any operator in $L[{\mcH}]$, then $T{{\bf a}}(\lambda )$ is of bounded variation and
\[
T\int_a^b {h(\lambda )} d{\mathbf{a}}(\lambda )=\int_a^b {h(\lambda )} dT{\mathbf{a}}(\lambda).
\]
\end{enumerate}
\end{thm}   
\subsection{Hilbert Space case}
with respect to a Hilbert space, our result shows that, in a well-defined sense, the Stone-von Neumann approach is generic.
\begin{thm} Let $A \in {\mcC}[{\mcH}]$ be arbitrary.  Then, for each $x \in D(A)$, there exists a vector-valued function ${\bf{e}}_x(\lambda )$ of bounded variation such that:
\begin{enumerate}
\item $D(A)$ also satisfies
\[
D(A) = \left\{ {\left. {x \in \mathcal{H}} \ \right|\;\int_{\sigma (A)} {\lambda ^2 \left({d{\bf{e}}_x(\lambda ),x} \right)_\mathcal{H}  < \infty } } \right\}
\]
and
\item 
\[
Ax = \mathop {\lim }\limits_{n \to \infty } \int_{ -n}^n {\lambda d{\bf{e}}_x(\lambda )}, \ {\rm for \ all} \; x \in D(A). 
\]
\item If $g(\cdot)$ is a complex-valued Borel function defined (a.e) on $\R$, then 
\[
g(A)= \mathop {\lim }\limits_{n \to \infty } \int_{ -n}^n {g(\lambda) d{\bf{e}}_x(\lambda )}  \ {\rm for \ all} \; x \in D_g(A),
\]
where
\[
D_g(A) = \left\{ {\left. {x \in \mathcal{H}} \ \right|\;\int_{\sigma (A)} {\lt|g(\lambda)\rt|^2 \left( {d{\bf{e}}_x(\lambda ),x} \right)_\mathcal{H}  < \infty } } \right\}.
\]
\end{enumerate}
\end{thm}
\begin{proof}  To prove (1), write $A=WR$, where $W$ is the unique partial isometry and $R=[A^{*}A]^{1/2}$.  By Theorem 5.1, there is a spectral measure ${\bf{E}}(\Om)$ such that, for each $x \in D(A)=D(R)$:   
\beqn
Rx= \mathop {\lim }\limits_{n \to \infty } \int_{ -n}^n {\lambda} d{\bf{E}}(d\lambda )x.
\eeqn
If we set ${\mathbf{a}}_x(\lambda )= {\bf{E}}(\lambda )x$, then ${\mathbf{a}}_x(\lambda )$ is a vector-valued function of bounded variation.  Furthermore, $W$ is a partial isometry and   $W{\mathbf{a}}_x(\lambda )$ is of bounded variation, with $Var(W{\mathbf{a}}_x,\R) \le Var({\mathbf{a}}_x,\R)$.  Thus, by Theorem 5.3, for each interval $(a,b)$, 
\[
W\int_a^b {\lambda} d{\mathbf{a}}_x(\lambda )=\int_a^b {\lambda} dW{\mathbf{a}}_x(\lambda ).
\]
Since $Ax=WRx$, if we set ${\mathbf{e}}_x(\lambda )=W{\mathbf{a}}_x(\lambda )$, we have from equation (5.1),   
\beqn
Ax= \mathop {\lim }\limits_{n \to \infty } \int_{-n}^n {\lambda} d{\mathbf{e}}_x(\lambda ).
\eeqn
The proof of (2) and (3) are now direct adaptations of the same result in \cite{DS}. 
\end{proof} 
Thus, with minor modification the the Stone-von Neumann Theorem extends to all closed densely defined linear operators on $\mcH$.
\begin{rem}
Given that $\{{\bf{E}}(\la) \}= \{ {\bf E}((-\iy, \la]) \}$ is any spectral family, Kato \cite{K} page 358, defined  $\lt |A \rt | =R=[A^*A]^{1/2}$ by:
\[
\lt |A \rt |x= \mathop {\lim }\limits_{n \to \infty } \int_{-n}^{n} \lt |{\lambda}\rt| d{\mathbf{{\bf{E}}}}(\lambda )x,  \ {\rm for} \; x \in D(A)= D(R). 
\]
This allowed him to show that different spectral families lead to different selfadjoint operators. (This result was also known to Stone \cite{SO} and von Neumann \cite{VN}.) 
 
It is clear that Theorem 5.4 could have been proven after 1948 when the book (Hille version) by Hille and Phillips appeared \cite{HP}.   It's a fact of history that, during the same period, research on vector-valued measures and abstract integration theory was also taking flight.  However, by this time, the interests of researchers in the field  had shifted to the use  of abstract methods in the study of operator algebras. This work led to a new version of the spectral theorem on Banach algebras via the well-known Gelfand transform (see Rudin \cite{RU}, Theorem 12.22, page 306).

We should point out that the disadvantage of Theorem 5.4 is that it gives no additional information at all about the known problems of spectral theory.   Thus, for the concrete problems of particular operators this is no help.  However, it does tell us that, for a given $A$, these problems are closely related to the detailed properties of the associated partial isometry $W$, with $A=WR$.  (Here, we presume that the properties of nonnegative selfadjoint operators are well understood?) 
\end{rem}
\subsection{Banach space case}
\begin{thm} If ${\mathcal{B}}' \subset {\mcH}_2$ and $A \in {\mcC}[{\mcB}]$ is the generator of a $C_0$-contraction semigroup, then there exists a unique vector-valued function ${\bf{e}}_x(\lambda )$ of bounded variation such that, for each $x \in D(A)$, we have:
\begin{enumerate}
\item $D(A)$ also satisfies  
\[
D(A) = \left\{ {\left. {x \in \mathcal{B}} \ \right|\;\int_{\sigma (A)} {\lambda ^2 \left< {d{\bf{e}}_x(\lambda ),f_x^s} \right>_\mathcal{B}  < \infty } } \right\}
\]
and
\item
\[
Ax= \mathop {\lim }\limits_{n \to \infty } \int_{ -n}^n {\lambda} d{\mathbf{e}}_x (\lambda ), \ {\rm for \ all} \; x \in D(A).
\]
\item If $g(\cdot)$ is a complex-valued Borel function defined (a.e) on $\R$, then $g(A) \in \mcC[\mcB]$.  Furthermore, 
\[
D_g(A) = \left\{ {\left. {x \in \mathcal{B}} \ \right|\;\int_{\sigma (A)} {\lt|g(\lambda)\rt|^2 \left< {d{\bf{e}}_x(\lambda ),f_x^s} \right>_\mathcal{B}  < \infty } } \right\}
\]
and
\item
\[
g(A)x= \mathop {\lim }\limits_{n \to \infty } \int_{ -n}^n {g(\lambda)} d{\mathbf{e}}_x (\lambda ),  \ {\rm for \ all} \; x \in D_g(A).
\]
\end{enumerate}
\end{thm}
\begin{proof}  By Theorem 3.4, $A=WR$, where $W$ is the unique partial isometry and $R=[A^{*}A]^{1/2}$.  Let $\bar R$ be the extension of $R$ to ${\mcH}_2$.  From equation (5.1), we see that  there is a unique spectral measure ${\bar {\bf E}}(\Om)$ such that for each $x \in D({\bar R})$:  
\beqn
{\bar R}x= \mathop {\lim }\limits_{n \to \infty } \int_{ 0}^n {\lambda} d{\bar {\bf E} }(d\lambda )x.
\eeqn
If we set ${\bar {\mathbf{a}}}_x(\lambda )= {\bar {\bf E}}(\lambda )x$, then ${\bar {\mathbf{a}}}_x(\lambda )$ is a vector-valued function of bounded variation.  Furthermore, if  $\bar W$ is the extension of $W, \   {\bar W}{\bar {\mathbf{a}}}_x(\lambda )$ is of bounded variation, with $Var({\bar W}{\bar {\mathbf{a}}}_x,\R) \le Var({\bar {\mathbf{a}}}_x,\R)$.  If we set ${\bar {\mathbf{e}}}_x(\lambda )={\bar W}{\bar {\mathbf{a}}}_x(\lambda )$, by Theorem 5.3, for each interval $(a,b)$, 
\[
\lt\{{\bar W}\int_a^b {\lambda} d{\bar {\mathbf{a}}}_x(\lambda )\rt\}=\int_a^b {\lambda} d{\bar {\mathbf{e}}}_x(\lambda ).
\]
Since ${\bar A}x={\bar W}{\bar R}x$ and the restriction of ${\bar A}$ to $\mcB$ is $A$, we have, for all $x \in D(A)$, 
\beqn
Ax= \mathop {\lim }\limits_{n \to \infty } \int_{-n}^n {\lambda} d{\mathbf{e}}_x(\lambda ).
\eeqn
This proves (2).  The proof of (1) follows from (1) in Theorem 5.1 and the definition of $f_x^s$.  The proofs of (3) and (4) are direct adaptations of the Hilbert space case (see \cite{RS}).
\end{proof}    
\subsection{General Case}
In this section, we assume that, for each $i, \ 1 \le i \le n, \ n \in \N, \ {\mcB}_i=\mcB$ is a fixed separable Banach space. We set $\mathfrak{B} =  \times _{i = 1}^n B_i $, and represent a vector $\bf x \in \mathfrak B$ by ${\mathbf{x}}^t  = \left[ {x_1 ,\;x_2 ,\; \cdots ,\;x_n } \right]$.  An operator ${\mathbf{A}} = \left[ {A_{ij} } \right] \in C[\mathfrak{B}]$ is defined whenever $A_{ij} :{\mcB} \to \mcB$,
 is in $\mcC [\mcB]$.
 
If ${\mcB}' \subset {\mcH}_2$ and $A_{ij}$ generates a $C_0$-contraction semigroup, then by Theorem 5.3, there exists a unique vector-valued function ${\bf{e}}_x^{ij}(\lambda )$ of bounded variation such that, for each $ x \in D(A_{ij})$, we have:
\begin{enumerate}
\item $D(A_{ij})$ also satisfies  
\[
D(A_{ij}) = \left\{ {\left. { x \in \mcB} \ \right|\;\int_{\sigma (A_{ij})} {\lambda ^2 \left< {d{\bf{e}}_x^{ij}(\lambda ),f_x^s} \right>_\mathcal{B}  < \infty } } \right\}
\]
and
\item
\[
A_{ij}x= \mathop {\lim }\limits_{n \to \infty } \int_{ -n}^n {\lambda} d{\mathbf{e}}_x^{ij} (\lambda ), \ {\rm for \ all} \; x \in D(A_{ij}).
\]
\item If $g(\cdot)$ is a complex-valued Borel function defined (a.e) on $\R$  then $g(A_{ij}) \in \mcC[\mcB]$.  Furthermore, 
\[
D_{g}(A_{ij}) = \left\{ {\left. {x \in \mathcal{B}} \ \right|\;\int_{\sigma (A_{ij})} {\lt|g(\lambda)\rt|^2 \left< {d{\bf{e}}_x^{ij}(\lambda ),f_x^s} \right>_\mathcal{B}  < \infty } } \right\}
\]
and
\item
\[
g(A_{ij})x= \mathop {\lim }\limits_{n \to \infty } \int_{ -n}^n {g(\lambda)} d{\mathbf{e}}_x^{ij} (\lambda ),  \ {\rm for \ all} \; x \in D_g(A_{ij}).
\]
\end{enumerate}
If we let  $d{\boldsymbol  {\mathcal E}}( {\la})=[d{\bf{e}}^{ij}(\la)]$,  then we can represent $\bf A$ and $g({\bf A})$ by: 
\[
{\bf A}{\bf x}= \mathop {\lim }\limits_{n \to \infty } \int_{ -n}^n { {\la}} d{\boldsymbol {\mathcal E}}( {\la}) {\bf x},  \ {\rm for \ all} \; {\bf x} \in D({\bf A})
\]
and 
\[
g({\bf A}){\bf x}= \mathop {\lim }\limits_{n \to \infty } \int_{ -n}^n g( {\la}) d{\boldsymbol {\mathcal E}}( {\la}) {\bf x},  \ {\rm for \ all} \; {\bf x} \in D({\bf A}).
\]

\section{{\bf Schatten Classes}} 
In this section, we show how our approach allows us to provide a natural definition for the Schatten class of operators on $\mathcal{B}$.

Let  $\mathbb{K}(\mathcal{B})$ be the class of compact operators on $\mathcal{B}$ and let $\mathbb{F}(\mathcal{B})$ be the set of operators of finite rank.  Recall that, for separable Banach spaces, $\mathbb{K}({\mathcal{B}})$ is an ideal that need not be the maximal ideal in $L[\mathcal{B}]$. If $\mathbb{M}(\mathcal{B})$ is the set of weakly compact operators and $\mathbb{N}(\mathcal{B})$ is the set of operators that map weakly convergent sequences into strongly convergent sequences, it is known that both are closed two-sided ideals in the operator norm, and, in general, $\mathbb{F}(\mathcal{B}) \subset \mathbb{K}(\mathcal{B}) \subset \mathbb{M}(\mathcal{B})$ and $\mathbb{F}(\mathcal{B}) \subset \mathbb{K}(\mathcal{B}) \subset \mathbb{N}(\mathcal{B})$ (see part I of Dunford and Schwartz \cite{DS}, pg. 553).  For reflexive Banach spaces,  $\mathbb{K}(\mathcal{B}) = \mathbb{N}(\mathcal{B})$ and $\mathbb{M}(\mathcal{B}) { = }L[\mathcal{B}]$.  For the space of continuous functions ${\mathbf{C}}[\Omega ]$, on a compact Hausdorff space $\Omega $, Grothendieck \cite{GO} has shown that $\mathbb{M}(\mathcal{B}) { = }\mathbb{N}(\mathcal{B})$.  On the other hand, it is shown in part I of Dunford and Schwartz \cite{DS} that, for a positive measure space, $\left( {\Omega ,\Sigma ,\mu } \right)$, on ${\mathbf{L}}^1 \left( {\Omega ,\Sigma ,\mu } \right), \; \mathbb{M}(\mathcal{B}) \subset \mathbb{N}(\mathcal{B})$.

We assume that ${\mathcal{B}}$ has the approximation property (i.e., every compact operator can be approximated by operators of finite rank).    (Recall, that ${\mathcal{H}}_1$ and ${\mathcal{H}}_2$ are fixed.)  Let $A$ be a compact operator on ${\mathcal{B}}$ and let $\bar A$ be its extension to $\mathcal{H}_2$.   For each compact operator $\bar A$ on $\mathcal{H}_2$, there exists an orthonormal set of functions $\{ \bar \varphi _n \,\left| {n \geqslant 1} \right.\} $ such that 
\[
\bar A = \sum\nolimits_{n = 1}^\infty  {\mu _n (\bar A)} \left( { \cdot \;,\bar \varphi _n } \right)_2 \bar U\bar \varphi _n,
\]
where the {$\mu _n $} are the eigenvalues of $[\bar A^*\bar A]^{1/2}  = \left| {\bar A} \right|$, counted by multiplicity and in decreasing order, and $\bar U$ is the partial isometry associated with the polar decomposition of $\bar A = \bar U\left| {\bar A} \right|$.  Without loss, we can assume that the set of functions $\{ \bar \varphi _n \,\left| {n \geqslant 1} \right.\} $ is contained in ${\mathcal{B}}$ and $\{ \varphi _n \,\left| {n \geqslant 1} \right.\} $ is the normalized version in ${\mathcal{B}}$.  If  $\mathbb{S}_p [\mathcal{H}_2 ]$ is the Schatten Class of order $p$ in $L[\mathcal{H}_2 ]$, it is well-known that, if $\bar A \in \mathbb{S}_p [\mathcal{H}_2 ]$, its norm can be represented as:
\[
\left\| {\bar A} \right\|_{_p }^{\mathcal{H}_2 }  = \{ Tr[\bar A^*\bar A]^{p/2} \} ^{1/p}  = \left\{ {\sum\nolimits_{n = 1}^\infty  {\left( {\bar A^*\bar A\bar \varphi _n ,\bar \varphi _n } \right)_{\mathcal{H}_2 }^{p/2} } } \right\}^{1/p}  = \left\{ {\sum\nolimits_{n = 1}^\infty  {\left| {\mu _n (\bar A)} \right|^p } } \right\}^{1/p}.
\]
\begin{Def} We represent the Schatten Class of order $p$ in $L[\mathcal{B}]$ by:
\[
\mathbb{S}_p [\mathcal{B}] = \mathbb{S}_p [\mathcal{H}_2 ] \cap L[\mathcal{B}]\left| {_\mathcal{B} } \right..
\]
\end{Def}
Since $\bar A$ is the extension of $A \in \mathbb{S}_p [\mathcal{B}]$, we can define $A$ on ${\mathcal{B}}$ by 
\[
A = \sum\nolimits_{n = 1}^\infty  {\mu _n (A)} \left\langle { \cdot \;,f_{n}^s (\varphi )} \right\rangle U\varphi _n, 
\]
where $f_{n}^s (\varphi )$ is the Steadman duality map associated with $\varphi _n $ and  $U$ is the restriction of  $ {\bar U}$ to ${\mathcal{B}}$.  The corresponding norm of $A$ on $\mathbb{S}_p [{\mathcal{B}}]$ is defined by: 
\[
\left\| A \right\|_{_p }^\mathcal{B}  = \left\{ {\sum\nolimits_{n = 1}^\infty  {\left\langle {A^*A\varphi _n ,f_{n}^s (\varphi )} \right\rangle ^{p/2} } } \right\}^{1/p}.
 \]

\begin{thm} Let $A \in \mathbb{S}_p [\mathcal{B}]$, then $\left\| A \right\|_{_p }^\mathcal{B}  = \left\| {\bar A} \right\|_{_p }^{\mathcal{H}_2 }$.
\end{thm}
\begin{proof} It is clear that $\{ \varphi _n \,\left| {n \geqslant 1} \right.\} $ is a set of eigenfunctions for $A^* A$ on ${\mathcal{B}}$.  Furthermore, by our extension of Lax's Theorem, $A^* A$ is selfadjoint and the point spectrum of $A^* A$ is unchanged by its extension to $\mathcal{H}_2 $.  It follows that $A^* A\varphi _n  = \left| {\mu _n} \right|^2 \varphi _n $, so that 
\[
\left\langle {A^* A\varphi _n ,f_{n }^s (\varphi )} \right\rangle  = \frac{\left| {\mu _n} \right|^2 }
{{\left\| {\varphi _n } \right\|_2^2 }}\left( { \varphi _n ,\varphi _n } \right)_2  = \left| {\mu_n } \right|^2, 
\]
and
\[
\left\| A \right\|_{_p }^\mathcal{B}  = \left\{ {\sum\nolimits_{n = 1}^\infty  {\left\langle {A^* A\varphi _n ,f_{n }^s (\varphi )} \right\rangle ^{p/2} } } \right\}^{1/p}  = \left\{ {\sum\nolimits_{n = 1}^\infty  {\left| {\mu _n } \right|^p } } \right\}^{1/p}  = \left\| {\bar A} \right\|_{_p }^{\mathcal{H}_2 }.
\]
\end{proof}

\begin{lem} If ${\mathcal{B}}$ has the approximation property, the embedding of $L[{\mathcal{B}}]$ in $L[\mathcal{H}_2 ]$ is both continuous and dense.
\end{lem}
\begin{proof} Recall that the embedding is continuous by Theorem \ref{L*: lax*}.   Since ${\mathcal{B}}$ has the approximation property, the finite rank operators $\mathbb{F}({\mathcal{B}})$ on ${\mathcal{B}}$ are dense in the finite rank operators $\mathbb{F}(\mathcal{H}_2)$ on $\mathcal{H}_2 $.  It follows that $\mathbb{S}_p [{\mathcal{B}}]$ is dense in $\mathbb{S}_p [\mathcal{H}_2 ]$.  In particular, $\mathbb{S}_1 [{\mathcal{B}}]$ is dense in $\mathbb{S}_1 [\mathcal{H}_2 ]$ and, since $\mathbb{S}_1 [\mathcal{H}_2 ]^*  = L[\mathcal{H}_2 ]$, we see that $\mathbb{S}_1 [{\mathcal{B}}]^*  = L[{\mathcal{B}}]$ must be dense in $L[{\mathcal{H}}_2 ]$. 
\end{proof}
It is clear that much of the theory of operator ideals on Hilbert spaces extend to separable Banach spaces in a straightforward way.  We state a few of the more important results to give a sense of the power provided by the existence of adjoints.  The first result extends theorems due to Weyl [WY], Horn [HO], Lalesco [LE] and Lidskii [LI].  (The methods of proof for Hilbert spaces carry over without much difficulty.)

\begin{thm} Let ${{A}} \in \mathbb{K}({\mathcal{B}})$, the set of compact operators on ${\mathcal{B}}$, and let $\{ {\lambda _n} \}$ be the eigenvalues of ${{A}}$ counted up to algebraic multiplicity.  If $\Phi $ is a mapping on $[0,\infty ]$ which is nonnegative and monotone increasing, then we have:
\begin{enumerate}
\item (Weyl)	
\[
\sum\nolimits_{n = 1}^{\mathbf{N}} {\Phi \left( {\left| {\lambda _n ({{A}})} \right|} \right)}  \leqslant \sum\nolimits_{n = 1}^{\mathbf{N}} {\Phi \left( {\mu _n ({{A}})} \right)} 
\]
and
\item (Horn)	
\[
\sum\nolimits_{n = 1}^{\mathbf{N}} {\Phi \left( {\left| {\lambda _n ({{A}}_1 {{A}}_2 )} \right|} \right)}  \leqslant \sum\nolimits_{n = 1}^{\mathbf{N}} {\Phi \left( {\mu _n ({{A}}_1 )\mu _n ({{A}}_2 )} \right)}. 
\]
In case ${{A}} \in \mathbb{S}_1 ({\mathcal{B}})$, we have:
\item (Lalesco)	
\[
\sum\nolimits_{n = 1}^{\mathbf{N}} {\left| {\lambda _n ({{A}})} \right|}  \leqslant \sum\nolimits_{n = 1}^{\mathbf{N}} {\mu _n ({{A}})} 
\]
and
\item (Lidskii)	
\[
\sum\nolimits_{n = 1}^{\mathbf{N}} {\lambda _n ({{A}})}  = Tr({{A}}).
\]
\end{enumerate}
\end{thm}
\subsection{{Discussion}}
In a Hilbert space ${\mathcal{H}}$, the Schatten classes $\mathbb{S}_p ({\mathcal{H}})$ are the only ideals in $\mathbb{K}({\mathcal{H}})$, and $\mathbb{S}_1 ({\mathcal{H}})$ is minimal.  In a Banach space, this is far from true.  A complete history of the subject can be found in the recent book by Pietsch \cite{PI1} (see also Retherford \cite{RE}, for a nice review).  We  limit this discussion to a few  major topics in the subject.  First, Grothendieck \cite{GO} defined an important class of nuclear operators as follows:

\begin{Def} If ${{A}} \in \mathbb{F}({\mathcal{B}})$ (the operators of finite rank), define the ideal ${\mathbf{N}}_1 ({\mathcal{B}})$
by:
\[
{\mathbf{N}}_1 ({\mathcal{B}}) = \left\{ {{{A}} \in \mathbb{F}({\mathcal{B}})\;\left| {\;{\mathbf{N}}_1 ({{A}}) < \infty } \right.} \right\},
\] 
where
\[
{\mathbf{N}}_1 ({{A}}) = \operatorname{glb} \left\{ {\sum\nolimits_{n = 1}^m {\left\| {f_n } \right\|\left\| {\phi _n } \right\|} \;\left| {f_n  \in {\mathcal{B}'},\;\phi _n  \in {\mathcal{B}},\;{{A}} = \sum\nolimits_{n = 1}^m {\phi _n \left\langle { \cdot \;,\,f_n } \right\rangle } } \right.} \right\}
\]
and the greatest lower bound is over all possible representations for ${{A}}$.
\end{Def}
Grothendieck has shown that ${\mathbf{N}}_1 ({\mathcal{B}})$ is the completion of the finite rank operators. ${\mathbf{N}}_1 ({\mathcal{B}})$ is a Banach space with norm ${\mathbf{N}}_1 ( \cdot )$, and is a two-sided ideal in $\mathbb{K}({\mathcal{B}})$.   It is easy to show that:

\begin{cor} $\mathbb{M}({\mathcal{B}}),\mathbb{N}({\mathcal{B}})$ and ${\mathbf{N}}_1 ({\mathcal{B}})$ are two-sided *ideals.
\end{cor}
In order to compensate for the (apparent) lack of an adjoint for Banach spaces, Pietsch \cite{PI2}, \cite{PI3}  defined a number of classes of operator ideals for a given ${\mathcal{B}}$.  Of particular importance for our discussion is the class $\mathbb{C}_p ({\mathcal{B}})$, defined by
\[
\mathbb{C}_p ({\mathcal{B}}) = \left\{ {{{A}} \in \mathbb{K}({\mathcal{B}})\;\left| {\,\mathbb{C}_p ({{A}}) = \sum\nolimits_{i = 1}^\infty  {[s_i ({{A}})]^p }  < \infty } \right.} \right\},
\]
where the singular numbers $s_n ({{A}})$ are defined by:
\[
s_n ({{A}}) = \inf \left\{ {\left\| {{{A}} - {{K}}} \right\|_{\mcB} \; \left| \ {{\text{rank of }}{{K}} \leqslant n} \right.} \right\}.
\]
Pietsch has shown that $\mathbb{C}_1 ({\mathcal{B}}) \subset {\mathbf{N}}_1 ({\mathcal{B}})$, while Johnson et al [JKMR] have shown that for each ${{A}} \in \mathbb{C}_1 ({\mathcal{B}})$, $\sum\nolimits_{n = 1}^\infty  {\left| {\lambda _n ({{A}})} \right|}  < \infty $.  On the other hand, Grothendieck [GO] has provided an example of an operator ${{A}}$ in ${\mathbf{N}}_1 (L^\infty  [0,1])$ with $\sum\nolimits_{n = 1}^\infty  {\left| {\lambda _n ({{A}})} \right|}  = \infty $ (see Simon [SI], pg. 118).   Thus, it follows that, in general, the containment is strict.  It is known that, if $\mathbb{C}_1 (\mathcal{B}) = {\mathbf{N}}_1 (\mathcal{B})$, then $\mathcal{B}$ is isomorphic to a Hilbert space (see Johnson et al).  It is clear from the above discussion, that: 

\begin{cor} $\mathbb{C}_p ({\mathcal{B}})$ is a two-sided *ideal in $\mathbb{K}({\mathcal{B}})$, and $\mathbb{S}_1 ({\mathcal{B}}) \subset {\mathbf{N}}_1 ({\mathcal{B}})$.
\end{cor}
For a given separable Banach space, it is not clear how the spaces $\mathbb{C}_p ({\mathcal{B}})$ of Pietsch relate to our Schatten Classes $\mathbb{S}_p ({\mathcal{B}})$ (clearly $\mathbb{S}_p ({\mathcal{B}}) \subseteq \mathbb{C}_p ({\mathcal{B}})$).  Thus, one question is that of the equality of $\mathbb{S}_p ({\mathcal{B}})$ and $\mathbb{C}_p ({\mathcal{B}})$.  (We suspect that  $\mathbb{S}_1 ({\mathcal{B}})=\mathbb{C}_1 ({\mathcal{B}})$.)
\section{Conclusion}
In this paper, we have refined and extended the work in \cite{GBZS} to develop a complete theory of adjoints for bounded linear operators on separable Banach spaces. We have further identified the obstacles to a similar program for closed densely defined linear operators.  A major result in this case is that all operators of Baire class one have an adjoint.  For applications, we restricted our consideration to generators of $C_0$-contraction semigroups.   We first used the polar decomposition property to extend the Poincar\'{e} inequality.   Then, the polar decomposition property, along with a few results for vector measures and vector-valued functions allowed us to extend the spectral theorem to all closed densely defined linear operators on separable Hilbert spaces.  Using our adjoint theory, we were able to  extend the spectral theorem to all bounded linear operators and all generators of $C_0$-contraction semigroups on separable Banach spaces.  As a final application, we introduced a new class of $^*$operator ideals on Banach spaces that parallel the Schatten class for Hilbert spaces.      
\acknowledgements
During the course of the development of this work, we have benefited from important critical remarks from Professor Ioan I. Vrabie.  

We would like to sincerely thank Professors Jerome Goldstein and Anatolij Pliczko for important correspondence on spectral operators and closed operators of Baire class on Banach spaces.  They also identified a few errors in an earlier draft, which led to an improvement in the paper.


\begin{thebibliography}{99}
\small
\bibitem[AL]{AL} A.  Alexiewicz,   { \it Linear functionals on Denjoy-integrable functions},  Colloq. Math. { \bf 1} (1948), 289-293.
\bibitem[ASV]{ASV} D. D. Ang, K. Schmitt and L. K. Vy, { \it A multidimensional analogue of the Denjoy-Perron-Henstock-Kurzweil integral}, Bull. Belg. Math. Soc. Simon Stevin {\bf 4} (1997), 355Ð371.
\bibitem[CF]{CF} I. Colojoar$\breve{a}$  and C. Foia\c{s}, { \it Theory of generalized spectral operators},  Gordon Breach, (1968).
\bibitem[DA]{DA} E. B. Davies, { \it The Functional Calculus},   J. London Mat. Soc. Vol. {\bf 52} (1995) 166-176.
\bibitem[DS]{DS} N. Dunford and  J. T.  Schwartz, { \it Linear Operators Part II: Spectral Theory},  Wiley Classics edition,  Wiley Interscience (1988).
\bibitem[DU]{DU} J. Diestel and J. J. Uhl, Jr, {\it Vector Measures },  Math. Surveys 15, Amer. Math. Soc. Providence, RI, (1977).
\bibitem[EV]{EV} L. C. Evans, { \it  Partial Differential Equations,}  AMS Graduate Studies in Math. {\bf 18},  Providence, R.I, 1998. 
\bibitem[GBZS]{GBZS} T.  Gill, S.  Basu,  W. W. Zachary and V. Steadman, { \it Adjoint for operators in Banach spaces}, Proceedings of the American Mathematical Society, { \bf 132} (2004), 1429-1434.
\bibitem[GO]{GO}\ A. \ Grothendieck, {\it Products tensoriels topologiques et espaces nucleaires}, Memoirs  of the American Mathematical Society, {\bf 16} (1955).
\bibitem[GR]{GR} L.  Gross, {\it Abstract Wiener spaces,} Proc. Fifth Berkeley Symposium on  Mathematics\  Statistics and Probability, (1965), 31-42.
\bibitem[GZ]{GZ} T. L. Gill and W. W. Zachary, {\it Foundations for relativistic quantum theory I: Feynman's operator calculus and the Dyson conjectures}, Journal of  Mathematical Physics  {\bf 43} (2002), 69-93.
\bibitem[GZ1]{GZ1} T. L. Gill and W. W. Zachary, {\it Banach Spaces for the Feynman integral}, Real Analysis Exchange  {\bf 34}(2) (2008)/(2009), 267-310.
\bibitem[GZ2]{GZ2} T. L. Gill and W. W. Zachary, {\it A New Class of Banach Spaces}, Journal of Physics A: Math. and Gen.  {\bf 41} (2008), 495206.
\bibitem [HO]{HO}\ A.\ Horn, {\it On the singular values of a product of completely continuous operators}, Proc.\ Nat.\ Acad.\ Sci.\ {\bf 36} (1950), 374--375.
\bibitem[HP]{HP} E. Hille and R. S. Phillips, {\it Functional Analysis and Semigroups}, Amer. Math. Soc. Colloq. Pub. 31, Amer. Math. Soc. Providence, RI, (1957).
\bibitem[HS]{HS} R. Henstock, {\it The General Theory of Integration}, Clarendon Press, Oxford, (1991).
\bibitem[HSJ]{HSJ} B. Helffer, and \ J.  Sj{\"o}strand,  {\it quation de Schrdinger avec champ magnetique et quation de Harper, Schrdinger Operators}, (Snderborg, 2988) eds. H. Holden and A Jensen,  Lecture Notes in Phys., vol. 345, Springer-Verlag, Berlin, (1989), 118--197.
\bibitem[JKMR]{JKMR}\ W. B. Johnson, \ H. Konig, \ B. Maurey and \ J. R. Retherford,  {\it Eigenvalues of p-summing and $l_p$ type operators in Banach space}, J. Funct. Anal. {\bf 32} (1978), 353--380.
\bibitem[K]{K} T. Kato,  { \it Perturbation Theory for Linear Operators,}  second ed. Springer-Verlag, New York, (1976).
\bibitem[KB]{KB}  J.  Kuelbs, {\it Gaussian measures on a Banach space,} Journal of Functional Analysis {\bf 5} (1970), 354--367.
\bibitem[KF]{KF} W.  E.  Kaufman, { \it A stronger metric for closed
operators in Hilbert spaces,} Proc. Amer. Math. Soc. {\bf 90} (1984), 83--87.
\bibitem[KPS]{KPS} S.G. Krein, Ju.I. Petunin and E.M. Semenov, {\it Interpolation of Linear Operators}, Nauka, Moscow, 1978; (English transl.), Translations Monographs 54, Amer. Math. Soc. Providence,  R.I. (1982).
\bibitem[KW]{KW} J.  Kurzweil, {\it Nichtabsolut konvergente Integrale}, Teubner-Texte z\"{u}r Mathematik, Band {\bf 26}, Teubner Verlagsgesellschaft, Leipzig, (1980).
\bibitem[LE]{LE} T.\  Lalesco, { \it Une theoreme sur les noyaux composes}, Bull. Acad. Sci. {\bf 3} (1914/15), 271--272.
\bibitem [LI]{LI} V.  B. Lidskii, { \it Non-self adjoint operators with a trace}, Dokl. Akad.  Nauk. SSSR {\bf 125} (1959), 485-487.
\bibitem [LP]{LP} G. Lumer and R. S. Phillips, Dissipative operators in a Banach space, {\it Pacific J. Math.} {\bf 11} (1961), 679-698.
\bibitem [LU]{LU} G. Lumer, Spectral operators, Hermitian operators and bounded groups, {\it Acta. Sci. Math.} (Szeged) {\bf 25} (1964), 75-85.
\bibitem[MO]{MO}P. Mikusi\'{n}ksi and K. Ostaszewski, { \it Embedding Henstock integrable functions into the space of Schwartz distributions}, Real Anal. Exchange {\bf 14}(1988-89), 24-29.
\bibitem[L]{L} P.  D.  Lax, Symmetrizable linear tranformations. \emph{Comm. Pure Appl. Math.} {\bf 7} (1954), 633--647.
\bibitem [PL]{PL} T. W. Palmer, Unbounded normal operators on Banach spaces, {\it Trans. Amer. Math. Sci.} {\bf 133} (1968), 385-414.
\bibitem[PF]{PF} W. F. Pfeffer, {\it The Riemann Approach to Integration: Local Geometric Theory}, Cambridge Tracts in Mathematics {\bf 109}, Cambridge University Press, (1993).
\bibitem[PI1]{PI1} A.  Pietsch, {\it History of Banach Spaces and Operator Theory}, Birkh\"{a}user, Boston, (2007).
\bibitem [PI2]{PI2} A.  Pietsch, {\it Einige neue Klassen von kompacter linear Abbildungen}, Revue der Math. Pures et Appl. (Bucharest), {\bf 8} (1963), 423--447.
\bibitem[PI3]{PI3}  A.  Pietsch, {\it Eigenvalues and s-Numbers },  Cambridge University Press, (1987).
\bibitem[PZ]{PZ} A. Pazy,  { \it Semigroups of Linear Operators and
Applications to Partial Differential  Equations} Applied Mathematical
Sciences,  {\bf 44}, Springer New York, (1983).
\bibitem[RE]{RE}\ J.\ R. Retherford, {\it Applications of Banach ideals of operators}, Bull. Amer. Math. Soc. {\bf 81} (1975), 978-1012.
\bibitem[RL]{RL} F. Rellich, {\it St\"{o}rungsterie der Spektralzerlegung V.},  Math. Ann.   {\bf 118} (1940), 462-484.
\bibitem[RS]{RS} M. Reed and B. Simon, {\it Methods of Modern Mathematical Physics I: Functional Analysis}, Academic Press, New York, (1972).
\bibitem[RU]{RU} W. Rudin, {\it Functional Analysis}, McGraw-Hill Press, New York, (1973).
\bibitem[SI]{SI}  B. Simon, {\it Trace Ideals and their Applications}, London Mathematical Society Lecture Notes Series {\bf 35}, Cambridge University Press, New York, (1979).
\bibitem[SL]{SL}  M. Slemrod, {\it Asymtotic behavior of $C_0$-semigroups as determined by the spectrum of the generator}, Indiana. Univ. Math. J. {\bf 25} (1976), 783-792. 
\bibitem[SO]{SO} M. H. Stone,  {\it Linear Transformations in Hilbert Space },  Math. Surveys 15, Amer. Math. Soc. Colloq. Publ. 15, Providence, RI, (1932).
\bibitem[ST]{ST}  V. Steadman, {\it Theory of operators on Banach spaces}, Ph.D thesis, Howard University, 1988.
\bibitem[STT]{STT} V. A. Strauss and C. Trunk, {\it Spectralizable Operators}, Integr. Equ. Oper. Theory {\bf 61} (2008), 413-422. 
\bibitem[TA]{TA} E. Talvila, { \it The distributional Denjoy integral}, Real Analysis Exchange {\bf 33} (2008), 51-82.
\bibitem[TI]{TI} E. C. Titchmarsh, {\it Some theorems on perturbation theory V.}, J. Analyse. Math. Soc. {\bf 4} (1954/56), 187-208. 
\bibitem [VN] {VN} J.\ von Neumann, {\it {\"U}ber adjungierte Funktionaloperatoren,} Annals of Mathematics {\bf 33} (1932), 294-310.
\bibitem[WY]{WY} H. Weyl, {\it Inequalities between the two kinds of  eigenvalues of a linear transformation}, Proc. Nat. Acad. Sci. {\bf 35},  (1949), 408-11.
\bibitem [VPP]{VPP} V. A. Vinokurov, Yu. Petunin and A. N. Pliczko, Measurability and Regularizability mappings inverse to continuous linear operators (in Russian), {\it Mat. Zametki.} {\bf 26} (1979), no. 4, 583-591.  English translation: { \it Math. Notes} {\bf 26} (1980), 781-785.
\bibitem[YS]{YS} K. Yosida,  { \it Functional Analysis,}  second ed. Springer-Verlag, New York, (1968).

\end{thebibliography}
\end{document}